\documentclass[10pt,journal,compsoc]{IEEEtran}
\usepackage{amsmath,graphicx,epsfig}
\usepackage[nocompress]{cite}
\usepackage{multirow}
\usepackage{booktabs}
\usepackage{float}
\usepackage{subfigure}
\usepackage{caption}
\usepackage{threeparttable}
\usepackage{diagbox}
\usepackage{url}
\usepackage{hyperref}
\hypersetup{
    colorlinks=true,
    linkcolor=black,
    filecolor=black,
    urlcolor=black,
    citecolor=black,
}
\usepackage{bm}
\usepackage{algorithm}
\usepackage{algorithmic}
\usepackage{bm}
\usepackage{amsfonts}

\usepackage[table,xcdraw]{xcolor}

\begin{document}
%
\title{LIQA: Lifelong Blind Image Quality Assessment}
%
%
%
%

\author{Jianzhao Liu, Wei Zhou,~\IEEEmembership{Student~Member,~IEEE}, Jiahua Xu, Xin Li, Shukun An and \\ Zhibo Chen,~\IEEEmembership{Senior~Member,~IEEE}
\thanks{This work was supported in part by NSFC under Grant U1908209, 61632001 and the National Key Research and Development Program of China 2018AAA0101400. (Jianzhao Liu and Wei Zhou contributed equally to this work.) (Corresponding anthor: Zhibo Chen.)}
\thanks{J. Liu, W. Zhou, J. Xu, X. Li, S. An and Z. Chen are with the CAS Key Laboratory of Technology in Geo-Spatial Information Processing and Application System, University of Science and Technology of China, Hefei 230027, China (e-mail: jianzhao@mail.ustc.edu.cn; weichou@mail.ustc.edu.cn; xujiahua@mail.ustc.edu.cn; lixin666@mail.ustc.edu.cn; ask@mail.ustc.edu.cn; chenzhibo@ustc.edu.cn).}
}

\markboth{Journal of \LaTeX\ Class Files,~Vol.~14, No.~8, April~2021}%
{Shell \MakeLowercase{\textit{et al.}}: Bare Demo of IEEEtran.cls for Computer Society Journals}

\IEEEtitleabstractindextext{%
\begin{abstract}
Existing blind image quality assessment (BIQA) methods are mostly designed in a disposable way and cannot evolve with unseen distortions adaptively, which greatly limits the deployment and application of BIQA models in real-world scenarios. To address this problem, we propose a novel Lifelong blind Image Quality Assessment (LIQA) approach, targeting to achieve the lifelong learning of BIQA. Without accessing to previous training data, our proposed LIQA can not only learn new distortions, but also mitigate the catastrophic forgetting of seen distortions. Specifically, we adopt the Split-and-Merge distillation strategy to train a single-head network that makes task-agnostic predictions. In the split stage, we first employ a distortion-specific generator to obtain the pseudo features of each seen distortion. Then, we use an auxiliary multi-head regression network to generate the predicted quality of each seen distortion. In the merge stage, we replay the pseudo features paired with pseudo labels to distill the knowledge of multiple heads, which can build the final regressed single head. Experimental results demonstrate that the proposed LIQA method can handle the continuous shifts of different distortion types and even datasets. More importantly, our LIQA model can achieve stable performance even if the task sequence is long.
\end{abstract}

}

\maketitle

\IEEEdisplaynontitleabstractindextext

%
\IEEEpeerreviewmaketitle

\IEEEraisesectionheading{\section{Introduction}\label{sec:introduction}}
\IEEEPARstart{B}{lind} image quality assessment (BIQA) is a challenging problem, which aims to automatically predict perceptual image quality without any information of reference images. It has received widespread attention due to the high demand in practical applications where reference images are difficult to obtain or even unavailable. Since various distortions would be generated at each stage of signal processing (e.g. acquisition, compression, and transmission), a reliable general-purpose BIQA algorithm is urgently needed. In addition, existing general-purpose BIQA models always suffer from catastrophic forgetting. Towards this end, the lifelong learning of BIQA which aims at tackling the catastrophic forgetting is significant in real-world applications.

Specifically, catastrophic forgetting \cite{mccloskey1989catastrophic} refers to the tendency that a neural network ``forgets" the previously learned knowledge upon sequentially learning new tasks. In IQA problem, we may encounter two settings which may bring the risk of catastrophic forgetting: distortion shift and dataset shift. Distortion shift occurs when we add a new distortion type to the current dataset. This is mainly because different distortion sensitivities of the human visual system (HVS) lead to distortion bias \cite{kim2017deep}. For example, for coding artifacts such as JPEG2000 and JPEG compression, lower sensitivities are assigned to image regions with higher activity while the artifacts in homogeneous regions are easier to observe. On the contrary, when the images are distorted by blur artifacts, strong edges are paid more attention rather than flat regions. Therefore, there exist differences in human perceptual judgments of different distortion types. That is, combining different distortions could result in perceptual conflicts which further causes catastrophic forgetting.

Dataset shift is a more general and complex situation beyond distortion shift. There are several factors that may contribute to the catastrophic forgetting of dataset shift. First, different datasets also have distinct distortion types in most cases. Second, a variety of perceptual scales are used in different datasets due to the differences in subjective testing methodologies \cite{zhou2018visual,xu2018subjective}. Images that have similar quality scores but are evaluated under different experimental settings may not be perceptually similar to each other. Third, the image contents of different datasets could vary widely. Images with similar perceptual scales but different contents may be assigned different quality scores by observers as a result of personal image content preferences \cite{gao2015learning}.  In this work, we investigate the LIQA model not only for the distortion shift, but also for the more complex dataset shift.

One straightforward way to mitigate catastrophic forgetting is joint training, which combines all training data. However, due to the storage limitation and privacy issues, previous data may not be accessible, making it hard to adopt the joint training strategy. In this paper, we focus on a more realistic and challenging setting of lifelong learning for BIQA which requires:
\begin{itemize}
    \item \textbf{Striking a balance between stability and plasticity.} Stability refers to the ability to preserve learned knowledge and plasticity denotes the fast adaptation to new knowledge. An ideal continual learner can achieve a stability-plasticity trade-off while learning new tasks.
    \item \textbf{Unavailability of the previous data source.} One cannot utilize the training data of previous tasks for joint training or pre-built memory sets to replay when a new task arrives.
    \item \textbf{A single-head architecture that provides task-agnostic predictions.} Multi-head architecture refers to allocating specific parameters to each task and needs task identity when testing. In the real world, we cannot accurately narrate distortion types and expect a deployed model to well suit for all tasks learned before.
    \item \textbf{Handling distortion shift and dataset shift.}  Task interference may occur when unseen distortion types or new datasets arrive. We expect a universal framework to handle both situations.
    \item \textbf{Robustness to task permutations.} Task order is an important factor because in some scenarios when the performance is impaired seriously by a certain task, the learning of subsequent tasks will be affected. We expect the model to resist the negative influence of certain tasks during the whole incremental learning process.
\end{itemize}

\begin{figure}[t]
\centering
\includegraphics[scale=0.45]{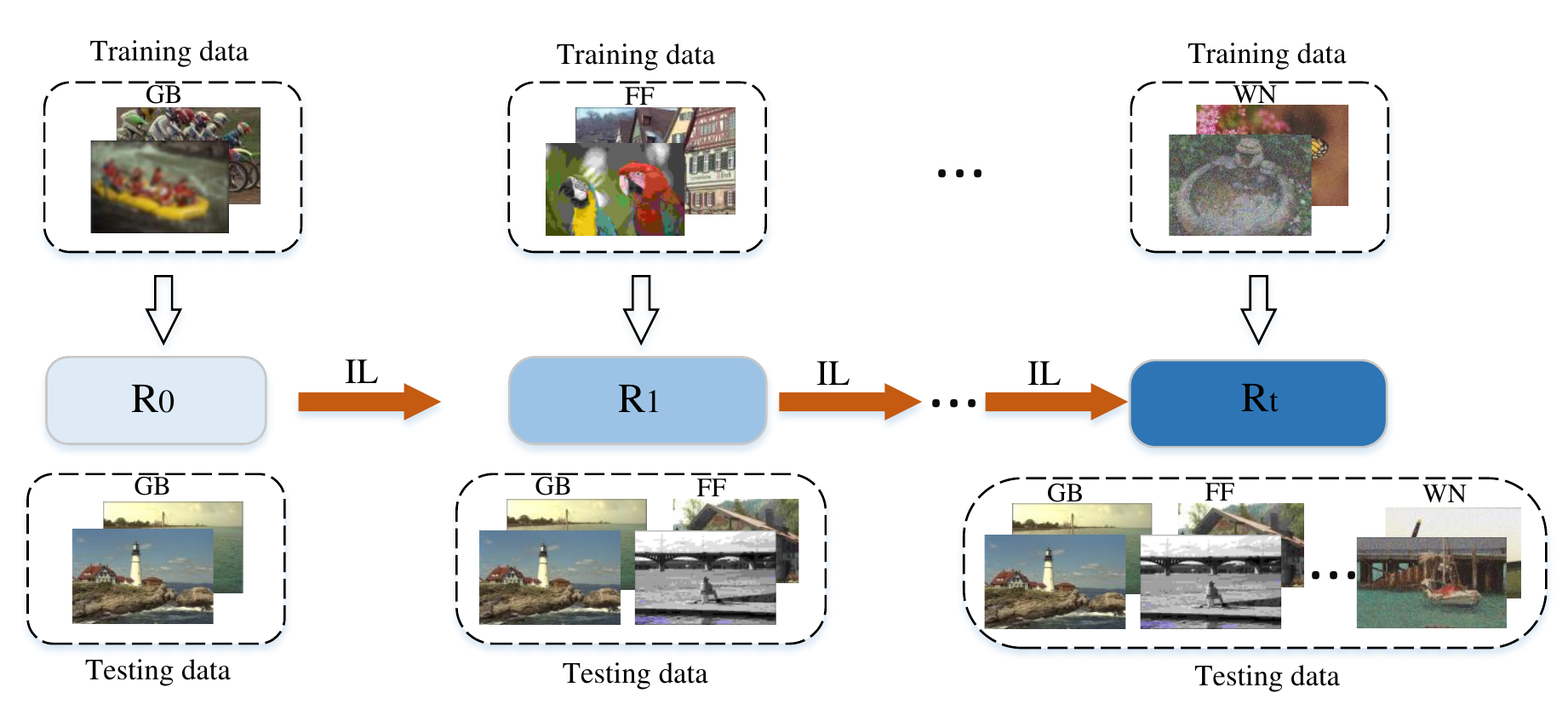}
\caption{Demonstration of the lifelong learning process of a BIQA network with new distortion types added sequentially.  $IL$ and $R$ denote incremental learning and regression network. The regression network needs to continually learn with the data arrival of new distortion types, without accessing to the training data of learned distortion types.}
\vspace{-0.5cm}
\label{fig:dist_shift}
\end{figure}

In Fig. \ref{fig:dist_shift}, we give an demonstration of the lifelong learning process of a BIQA network with new distortion types added sequentially. Without accessing to the training data of learned distortion types, the network continually evolves with the data arrival of new distortion types.

In this paper, we propose a new Lifelong blind Image Quality Assessment (LIQA) model aiming to tackle the catastrophic forgetting of BQIA. Unlike most classification methods that directly distillate knowledge from the former network, we adopt the Split-and-Merge strategy to train a single-head regression network. In the split stage, we first employ a generator with distortion-specific heads to memorize the data distributions of each seen distortion type. We split the regression network into a feature extractor and a prediction head. Instead of generating pseudo images, we choose to generate pseudo features before the prediction head. The features are compact image representations and also the direct inputs of the regression head. Besides, features have lower dimensions compared with images (e.g. a $256 \times 256$ image can be embedded into a $512$-dimensional vector) and are easier to be generated when the training data is limited. The generator is conditioned on the distortion type and the quality score. It can well control the category of pseudo features, thus avoiding the potential unbalanced problem of distortion types and quality ranges. Once the generator is trained, it can serve as a memory replayer to replay the generated pseudo features which resemble the real features of previous distortions. Then, we utilize an auxiliary multi-head regression network to  generate pseudo labels with respect to the pseudo features of each distortion. In the merge stage, we distill the knowledge of the auxiliary multi-head regression network using the pseudo features and the corresponding pseudo labels to build the single-head regression network. It should be noted that the Split-and-Merge strategy effectively resists the negative impact of a certain task during the sequential learning process and improves the robustness to task permutations. Suppose that we distill knowledge from the previous single-head network instead of the auxiliary multi-head network. Some previous distortions' pseudo labels are probably inaccurate due to forgetting at the completion of the previous task. When learning a new task, this will cause the continuous performance drop of these distortions due to error propagation, which is especially harmful when the task sequence is long.

In summary, our contributions are as follows:
\begin{itemize}
    \item We drill down into the lifelong learning of BIQA and propose a LIQA framework which can effectively mitigate the catastrophic forgetting when learning new tasks.
     \item We adopt the Split-and-Merge strategy to train a single-head regression network, which can resist the negative effects of certain tasks and improve robustness to task permutations.
    \item We design a generator that well controls the generation of pseudo features conditioned on the distortion type and the quality score. It serves as a memory replayer to consolidate the learned knowledge while learning new task, avoiding the error caused by the unbalanced distribution of distortion types and quality ranges.
    \item We conduct experiments to verify the effectiveness of LIQA when meeting with the continuous shifts of different distortion types and datasets.
\end{itemize}

The remaining parts of this paper are organized as: We review related work in Section \ref{sec:related work}. In Section \ref{sec:preliminary}, we introduce background knowledge. Our approach is presented in Section \ref{sec:approach} and experimental results are reported in Section \ref{sec:experiments}. We conclude our work and discuss several future directions in Section 6. Our codes are available to the research community at \url{http://staff.ustc.edu.cn/~chenzhibo/resources.html}.

\section{Related works}
\label{sec:related work}
\subsection{Blind image quality assessment}
BIQA aims to automatically predict the subjective quality, i.e. the mean opinion score (MOS) or differential mean opinion score (DMOS), of a distorted image without accessing to the reference information. It can be roughly divided into two categories: distortion-specific and general-purpose approaches \cite{BIQA_sur}. Distortion-specific methods are designed for a particular distortion type (e.g. JPEG \cite{JPEG-BIQA}, blur \cite{BLUR-BIQA} and super-resolution \cite{SR-BIQA}). These methods deliver a poor generalization ability to other distortion types and can only be tested when the distortion type is known. In contrast, general-purpose methods that can perform across various distortion types are more practical. Natural images show a number of consistent statistical properties that have been explored in spatial or transform domain in many research works. BRISQUE \cite{mittal2012no} utilizes statistics measured in the spatial domain and employs a generalized Gaussian distribution (GGD) model to capture various distorted image statistics.
BLIINDS \cite{dct} predicts image quality according to DCT-based contrast and DCT-based structural features. Apart from the above-mentioned handcrafted features, there are many learning-based BIQA methods \cite{zhou2019dual,xu2020blind} that usually follow a two-step network, i.e.,  feature extraction and quality prediction. Pan et al. \cite{pan2018blind} employed a fully convolutional neural network to predict a pixel-by-pixel similar quality map and used a deep pooling network to regress the quality map into a score. In \cite{zhang2018blind}, Zhang et al. proposed a deep bilinear model that works for both synthetically and authentically distorted images. Recently, improving the training strategy of BIQA has become popular. Gao et al. \cite{gao2015learning} exploited preference image pairs to address the problem of insufficient training data. In \cite{zhang2020learning}, Zhang et al. learned data uncertainty and trained a deep neural network over massive image pairs by minimizing the fidelity loss. In \cite{zhang2021uncertainty}, Zhang et al. further used the uncertainty training strategy and developed UNIQUE, which can obtain better generalization ability in the cross-database setting.

Although these BIQA methods have achieved great success, they are designed in a disposable way and lack the ability to evolve with unseen distortions. In contrast, our work tries to explore the sustainable learning ability of BIQA networks and thus helps BIQA networks to accumulate new knowledge and retaining the learned knowledge at the same time.

\subsection{Lifelong learning}
Lifelong learning is also referred to as incremental learning or continual learning. The major challenge for lifelong learning is the stability-plasticity dilemma \cite{grossberg2012studies},  where plasticity represents the ability to integrate new knowledge and stability requires that performance on previously learned tasks should not significantly degrade over time as new tasks are added. Lifelong learning methods can be roughly divided into three categories: regularization methods, parameter isolation methods \cite{delange2021continual} and replay methods. Prior-focused regularization-based approaches, such as EWC \cite{ewc}, online EWC \cite{onlineewc} and SI \cite{si}, usually add a regularization term that discourages the alteration to weights important to previous tasks, which effectively prevents old knowledge from being erased or overwritten. Data-focused regularization-based methods, such as LWF \cite{lwf}, LFL \cite{jung2016less} and DMC \cite{zhang2020class}, employ a  distillation loss to encourage the responses to previous tasks remain unchanged. Parameter isolation methods allocate task-specific parameters. One can dynamically accommodate new branches while freezing previous task parameters if there are no constraints on network size  \cite{rusu2016progressive,xu2018reinforced,rosenfeld2018incremental}. When the architecture remains static, parameters of fixed parts are allocated to different tasks. HAT \cite{HAT} learns hard attention masks to each task at the unit level. PackNet \cite{mallya2018packnet} iteratively assigns parameter subsets to consecutive tasks by constituting binary masks. Replay methods such as iCaRL \cite{rebuffi2017icarl} and ER \cite{rolnick2018experience}  use representative samples selected from the small memory set while learning a new task. Due to the storage and privacy issues, the previous training data may be unavailable. Therefore, some replay methods such as DGR \cite{DGR}, PR \cite{PR}, GFR \cite{GFR} and BIR \cite{van2020brain} utilize Generative Adversarial Network (GAN) to generate pseudo images or features to consolidate the learned knowledge.

In spite of the impressive successes in the filed of classification, it still remains a challenge for the BQIA due to the particularity of quality assessment. First, the size of IQA datasets is limited. Second, the quality label of IQA is continuous instead of discrete like classification. It lacks the obvious boundaries of data due to the low aggregation of data especially for the authentically distorted images. In this case, we take steps towards the lifelong learning of BIQA and try to find a suitable way for BIQA networks to continually learn.

\section{Preliminaries}
\label{sec:preliminary}
\subsection{Problem definition}
We first define a task sequence $\mathbf{T}=\{T_{t}\}_{t=0}^{N}$,  where  $N+1$ denotes the total number of tasks. Each task $T_{t}$ consists of a set of new distortion types. During learning current task $T_{\mathcal{T}}$, we can only get accesss to training data $D_{\mathcal{T}}=(\mathcal{X}^\mathcal{T},\mathcal{S}^\mathcal{T})=\{(\mathbf{x}_{i}^{\mathcal{T}},{s}_{i}^{\mathcal{T}})\}_{i=1}^{{n}_{\mathcal{T}}}$, where $\mathbf{x}_{i}^{\mathcal{T}}$ represents the distorted image,  ${s}_{i}^{\mathcal{T}}$ represents the ground truth perceptual quality score and ${n}_{\mathcal{T}}$ denotes the number of the $\mathcal{T}$-th task's training data. For any $t_{1} \neq t_{2}$, $D_{t_{1}}\cap D_{t_{2}}=\emptyset$. The ultimate goal of current task $T_{\mathcal{T}}$ is to control the statistical risk of all seen tasks given no access to data from previous tasks $T_{t<\mathcal{T}}$. $T_{0}$ is denoted as a base task, which consists of $M_{0}$ distortion types. Supposing the total number of  distortion types as $M_{all}$, we can sequentially add  $\Delta = \frac{M_{all}-M_{0}}{N}$ distortion types per task, resulting in $N$ novel tasks (i.e. $T_{1}-T_{N}$). We can say that the incremental step is equal to $\Delta$. After the  model has incrementally been trained up to $\mathcal{T}$-th task ($\mathcal{T}>0$), we denote $M_{cur}={{M}_{0}}+\Delta*\mathcal{T}$ as the number of distortion types seen so far and denote  $M_{pre} = {{M}_{0}}+\Delta*(\mathcal{T}-1)$ as the number of all previously seen distortions before learning the current task.

\subsection{Evaluation metrics}
For all experiments, we specially design two evaluation metrics for lifelong learning of BIQA: Correlation Index (C) and Forgetting Index (F), following the previous works \cite{chaudhry2018riemannian,diaz2018don}.

\textbf{Correlation Index (C)}
After learning the $\mathcal{T}$-th task, we evaluate the average Spearman's Rank-order Correlation Coefficient (SRCC) between the predicted quality scores and the MOS/DMOS on the held-out test images of each seen distortion type. The  correlation index $C_{\mathcal{T}}$ is defined as $C_{\mathcal{T}}=\frac{1}{M_{cur}}\sum\limits_{j=0}^{M_{cur}-1}abs(\text{SRC}{{\text{C}}_{\mathcal{T},j}})$,
 which is within the range of [0,1]. The higher value means the better consistency with human opinions of perceptual quality.

\textbf{Forgetting Index (F)}
We define the forgetting degree of a particular distortion as the difference between the maximum performance of this distortion through out the learning process in  the past and the performance the current model has about it.   The forgetting of the $j$-th distortion after the model has incrementally been trained up to $\mathcal{T}$-th task can be defined as:
\begin{equation}
    f_{j}^{\mathcal{T}}=\underset{t\in \{0,\ldots ,\mathcal{T}-1\}}{\mathop{\max }}\,abs(\text{SRC}{{\text{C}}_{t,j}})-abs(\text{SRC}{{\text{C}}_{\mathcal{T},j}})
\end{equation}
where $\mathcal{T}>0$, $j \in [0, M_{pre})$ and  $f_{j}^{\mathcal{T}} \in [-1,1]$. $abs(\cdot)$ denotes the absolute value function. We can average the forgetting of all previously seen distortions and obtain the forgetting index $F_{\mathcal{T}}=\frac{1}{M_{pre}}\sum\limits_{j=0}^{M_{pre}-1}{f_{j}^{\mathcal{T}}}$. Lower $F_{\mathcal{T}}$ means less forgetting and better stability. Especially, when $F_{\mathcal{T}}<0$, it means that the current task not only cannot impair the previous learned knowledge but also can contribute to the performance of previous tasks.

\section{Approach}
\label{sec:approach}

\subsection{Network architecture of LIQA}
The framework of LIQA consists of four parts: a single-head regression network $\mathcal{R}$, an auxiliary multi-head regression network $\hat{\mathcal{R}}$, a generator $\mathcal{G}$ and a discriminator $\mathcal{D}$. For the single-head regression network, we employ a pre-trained ResNet-18 (without the final $FC$ layers) as the feature extractor $U$ and use two $FC-ReLU$ layers followed by a $Sigmoid$ function as the prediction head $V$. The auxiliary multi-head regression network has distortion-specific prediction heads $\hat{V}_{j=0}^{{{M}_all}-1}$ and the architecture of the feature extractor $\hat{U}$ is the same as that of the single-head regression network's.

\begin{figure}[htbp]
\centering
\includegraphics[scale=0.38]{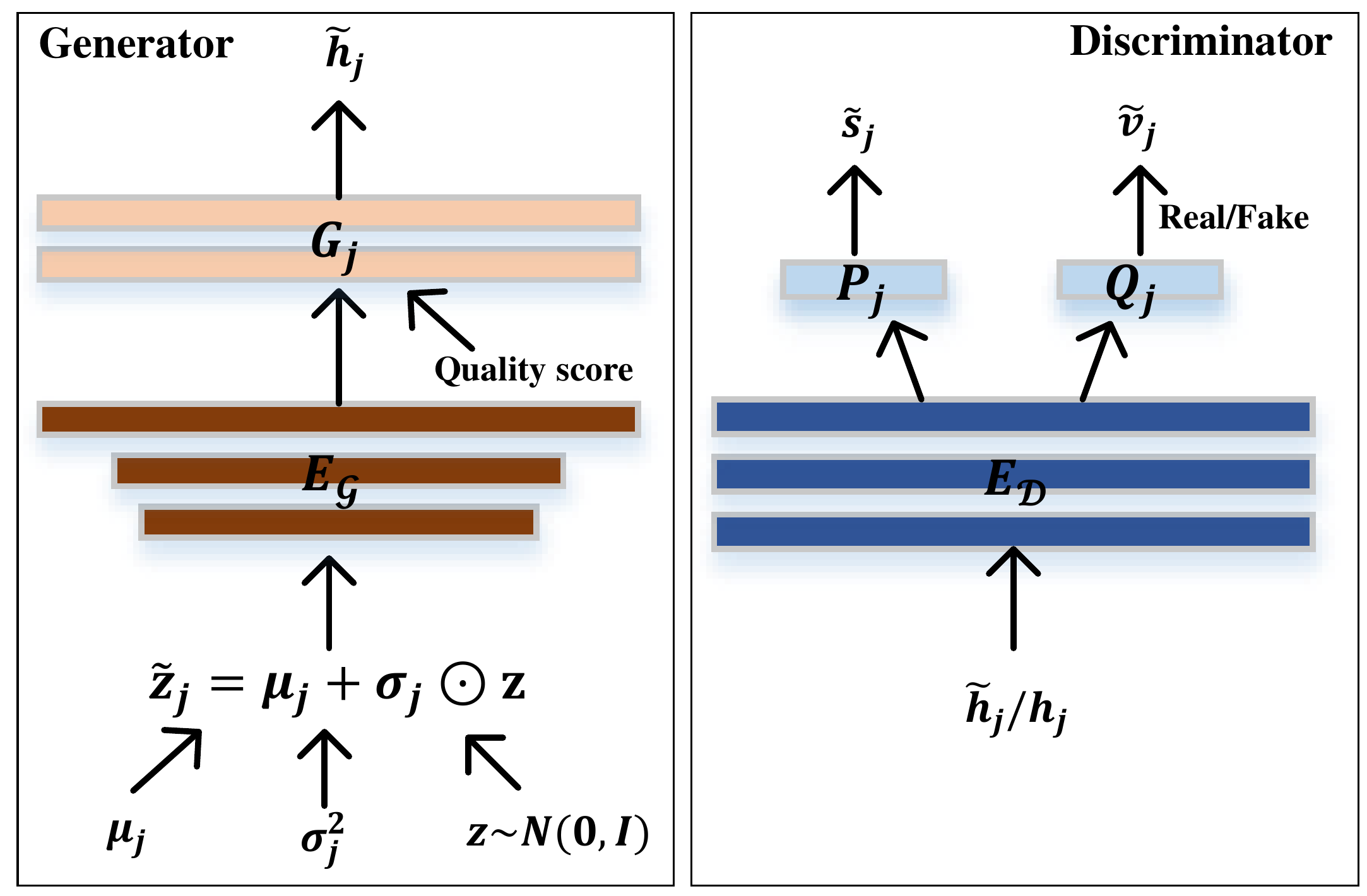}
\caption{Architectures of the generator and the discriminator.}
\label{fig:G_And_D}
\end{figure}

The architectures of the generator and the discriminator are shown at Fig. \ref{fig:G_And_D}. Instead of sampling noise vector form the standard normal prior $\mathcal{N}(\mathbf{0},\mathbf{I})$, we sample noise vector $\Tilde{\mathbf{z}}_{j}$ from $\mathcal{N}(\bm{\mu}_{j},\bm{\sigma}_{j}^2)$, where $\bm{\mu}_{j}$ and $\bm{\sigma}_{j}$ are trainable mean and standard deviation for the distortion $j$.  We adopt the reparameterization trick \cite{VAE} to generate $\Tilde{\mathbf{z}}_{j}$. $\Tilde{\mathbf{z}}_{j} = \bm{\mu}_{j} + \bm{\sigma}_{j} \odot \mathbf{z}$, where $\mathbf{z} \sim \mathcal{N}(\mathbf{0},\mathbf{I})$ and $\odot$ signifies the element-wise product. This makes it possible to generate distortion-specific features by restricting the sampling of the noise vector from the corresponding distribution. The generator consists of two parts: the shared embedding layers $E_{\mathcal{G}}$ which embed the noise vector to a latent vector and the distortion-specific generation head $G_{j}$ which takes in the summation of the latent vector and the quality score to generate quality-related distortion-specific pseudo features $\tilde{\mathbf{h}}_{j}$.

The discriminator consists of three parts: the shared embedding layers $E_{\mathcal{D}}$ which embed the input feature vector to a latent vector, the distortion-specific quality prediction heads $P_{j}$ which regresses the latent vector into a quality score $\Tilde{s}_{j}$and the distortion-specific discrimination head $Q_{j}$ which tells whether the input feature is real or fake.

Generally speaking, the generator is conditioned on the
 distortion index  $j$ and the quality score $s$, which can be denoted by $\tilde{\mathbf{h}}_{j}=\mathcal{G}(\mathbf{z},s,j)$. The discriminator takes in the pseudo/real feature $\Tilde{\mathbf{h}}_{j}/\mathbf{h}_{j}$ together with the distortion index $j$, which can be denoted by $(\Tilde{s}_{j},\Tilde{v}_{j})=\mathcal{D}(\Tilde{\mathbf{h}}_{j}/\mathbf{h}_{j},j)$.

\begin{figure*}[htbp]
\centering
\includegraphics[scale=0.48]{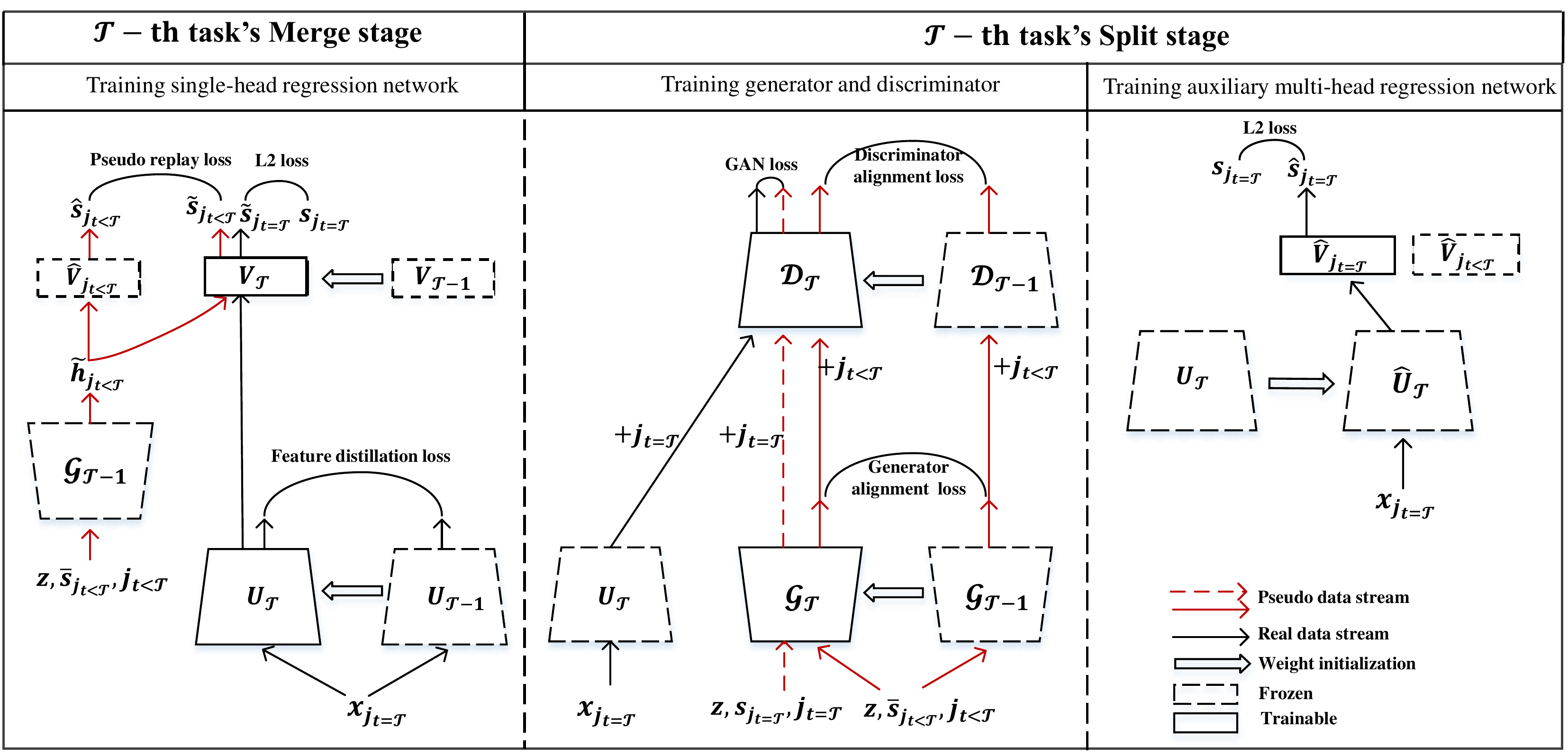}
\caption{Framework of LIQA. LIQA adopts Split-and-Merge distillation strategy to build the single-head regression network.
$U$ and $\hat{U}$ are the feature extractor of the single-head regression network and the multi-head regression network.
$V$ and $\hat{V}$ are the prediction head of the single-head regression network and the multi-head regression network. $\mathcal{G}$ is the generator and  $\mathcal{D}$ is the discriminator. $\mathbf{z}$, $s$ and $j$ denotes the random noise vector, the quality score and the distortion index respectively.   }
\label{fig:framework of LIQA}
\end{figure*}

\subsection{Training strategy}
The whole framework of LIQA is shown in Fig. \ref{fig:framework of LIQA}. In the merge stage, we train the single-head regression network with the pseudo features replaying. In the split stage, we incrementally train the generator and the discriminator and then train the auxiliary multi-head regression network to learn each seen distortion type separately.

\subsubsection{Training single-head regression network}
Let us denote the current task by $T_{\mathcal{T}}$. The generator and the discriminator trained at the split stage at the former task is $\mathcal{G}_{\mathcal{T}-1}$ and $\mathcal{D}_{\mathcal{T}-1}$. The feature extractor and the prediction head of the current single-head network $\mathcal{R}_{\mathcal{T}}$ is $U_{\mathcal{T}}$ and  $V_{\mathcal{T}}$ respectively. The prediction heads of the auxiliary multi-head regression network trained at previous tasks are $\hat{V}_{j_{t<{\mathcal{T}}}}$, where the subscript $j_{t<{\mathcal{T}}}$ denotes the index of the distortions at task  $\mathcal{T}_{t<\mathcal{T}}$.

\textbf{Training the feature extractor.} Instead of freezing some layers of the feature extractor like \cite{jung2016less}, we fine-tune all the parameters of the feature extractor, improving the plasticity of the network and leaving more room for new task learning. Moreover, we employ feature distillation loss  to prevent the forgetting of old knowledge and guarantee the stability of the feature extractor.
 The feature distillation loss during task $T_{\mathcal{T}}$ is defined as:

\begin{equation}
\centering
    L_{\mathcal{T}}^{FD}={{\mathbb{E}}_{\mathbf{x}\sim {{\mathcal{X}}^{\mathcal{T}}}}}\left[{{\left\| {{U}_{\mathcal{T}}}(\mathbf{x})-{{U}_{\mathcal{T}-1}}(\mathbf{x}) \right\|}_{2}}\right],
\label{equ:FD}
\end{equation}
which is used to constrain that the features extracted by ${U}_{\mathcal{T}}$ do not drift far away from that by ${U}_{\mathcal{T}-1}$.

\textbf{Training the prediction head.} For training the prediction head, we employ pseudo replay loss to consolidate the knowledge of previous distortions and use $L_{2}$ loss to learn the new distortions arriving at $T_{\mathcal{T}}$. Given a random quality score $\bar{s}$, a random noise $\mathbf{z}$ and the distortion index ${j}_{t<\mathcal{T}}$, ${\mathcal{G}}_{\mathcal{T}-1}$ can generate the pseudo quality-related feature of distortion ${{j}_{t<\mathcal{T}}}$: ${{\Tilde{\mathbf{h}}}_{{{j}_{t<\mathcal{T}}}}}=\mathcal{G}_{\mathcal{T}-1}(\mathbf{z},\bar{s},{{j}_{t<\mathcal{T}}})$. The pseudo replay loss can be defined as:
\begin{equation}
\begin{aligned}
\label{equ:pr}
{{L}^{PR}}={{\mathbb{E}}_{z\sim \mathcal{N}(\mathbf{0},\mathbf{1}),s \sim \bar{\mathcal{S}},j \sim p_{j_{t<\mathcal{T}}} }}\left[{{\left\| {{{\hat{V}}}_{j}}({{{\tilde{h}}}_{j}})-{{V}_{\mathcal{T}}}({{{\tilde{h}}}_{j}}) \right\|}_{2}}\right],
    \end{aligned}
\end{equation}
where ${{{\tilde{h}}}_{j}} = \mathcal{G}_{\mathcal{T}-1}(\mathbf{z},s,j)$. $s \sim \bar{\mathcal{S}}$ means the quality  score is randomly generated and $p_{j_{t<\mathcal{T}}}$ is the distortion index distribution of previous tasks $T_{t<{\mathcal{T}}}$.  ${{\hat{V}}_{j}}$ is the prediction head for distortion ${j}$ of the auxiliary multi-head regression network. For current task's training data, we adopt $L_{2}$ loss for training:
\begin{equation}
  L_{\mathcal{T}}^{MSE1}={{\mathbb{E}}_{(\mathbf{x},s)\sim {{D}^{\mathcal{T}}}}}\left[ {{\left\| {{V}_{\mathcal{T}}}({{U}_{\mathcal{T}}}(\mathbf{x}))-s \right\|}_{2}} \right]
\end{equation}

\textbf{Full objective.} In summary, when $\mathcal{T}=0$, we only have $L_{2}$ loss for training the single-head regression network:
\begin{equation}
     L_{\mathcal{T}}^{Total} = L_{\mathcal{T}}^{MSE1}.
\end{equation}
During task $T_{\mathcal{T}}$ ($\mathcal{T}>0$), the total training loss is:
\begin{equation}
    L_{\mathcal{T}}^{Total}={{\lambda }_{FD}}L_{\mathcal{T}}^{FD}+{{\lambda }_{PR}}L_{\mathcal{T}}^{PR}+{{\lambda }_{MSE}}L_{\mathcal{T}}^{MSE1},
    \label{equ: full_single}
\end{equation}
where ${\lambda }_{FD}$, ${\lambda }_{PR}$ and ${\lambda }_{MSE}$ are hyper-paramters that control the relative importance of feature distillation loss, pseudo replay loss and $L_{2}$ loss respectively.

\subsubsection{Training generator and discriminator}
We freeze the single-head network trained at task $T_{\mathcal{T}}$ and train the generator $\mathcal{G}_{\mathcal{T}}$ to continually learn the current task's feature distribution. It should be noted that the trainable ${\bm{\mu}_{j_{t<\mathcal{T}}}}$,  ${\bm{\sigma}_{j_{t<\mathcal{T}}}}$ and the previously learned generation heads ${G}_{j_{t<\mathcal{T}}}$ of $\mathcal{G}_{\mathcal{T}}$ are also frozen. Similarly, the previous quality prediction heads ${P_{j_{t<\mathcal{T}}}}$ and discrimination head ${Q_{j_{t<\mathcal{T}}}}$ of ${\mathcal{G}_{\mathcal{T}}}$ are frozen.

\textbf{Adversarial loss.} To make the generated pseudo features indistinguishable from real features, we adopt an adversarial loss:
\begin{small}
\begin{equation}
\begin{aligned}
   L_{\mathcal{T}}^{adv}=&{{\mathbb{E}}_{\mathbf{x}\sim {{\mathcal{X}}^{\mathcal{T}}},j \sim p_{j_{t=\mathcal{T}}}}}\left[ \log {{\mathcal{D}}_{\mathcal{T}}^{r/f}}({{U}_{\mathcal{T}}}(\mathbf{x}),j)) \right]+\\&{{\mathbb{E}}_{{\mathbf{z}\sim \mathcal{N}(\mathbf{0},\mathbf{I})},s \sim {{\mathcal{S}}}^{\mathcal{T}},j \sim p_{j_{t=\mathcal{T}}}}}\left[ \log (1-{{\mathcal{D}}_{\mathcal{T}}^{r/f}}({{\mathcal{G}}_{\mathcal{T}}}(\mathbf{z},s,j),j)) \right]
 \end{aligned}
\end{equation}
\end{small}
where ${{\mathcal{D}}_{\mathcal{T}}^{r/f}}$ is the discrimination head for distortion $j$ which gives the real/fake probability. $\mathcal{G}_{\mathcal{T}}$ tries to generate a feature conditioned on both the quality score $s$ and the distortion index $j$, while   $\mathcal{D}_{\mathcal{T}}$ tries to distinguish between the real and the fake features of distortion $j$.

\textbf{Quality prediction loss.} To generate pseudo features corresponding to the quality score $s$, we add auxiliary quality prediction heads on top of the discriminator and impose the quality prediction loss when optimizing both $\mathcal{D_\mathcal{T}}$ and $\mathcal{G_{\mathcal{T}}}$: a quality prediction loss of real features used to optimize $\mathcal{D_\mathcal{T}}$, and a quality prediction loss of fake features used to optimize $\mathcal{G_{\mathcal{T}}}$. In detail, the former is defined as
\begin{equation}
    L_{\mathcal{T}}^{qua\_r}={{\mathbb{E}}_{(\mathbf{x},s)\sim {{D}^{\mathcal{T}}},j \sim p_{j_{t=\mathcal{T}}}}}\left[ {{\left\| {{\mathcal{D}}_{\mathcal{T}}^{qua}}({{U}_{\mathcal{T}}}(\mathbf{x}),j)),s \right\|}_{2}} \right],
\end{equation}
where the term ${{\mathcal{D}}_{\mathcal{T}}^{qua}}$ represents the the quality prediction head which predicts the quality value. On the other hand, the quality prediction loss of fake features is defined as
\begin{small}
\begin{equation}
    L_{\mathcal{T}}^{qua\_f}={{\mathbb{E}}_{\mathbf{z}\sim \mathcal{N}(\mathbf{0},\mathbf{I}),s \sim \mathcal{S}^{\mathcal{T}},j \sim p_{j_{t=\mathcal{T}}}}}\left[ {{\left\| {{\mathcal{D}}_{\mathcal{T}}^{qua}}({{\mathcal{G}}_{\mathcal{T}}}(\mathbf{z},s,j),j),s \right\|}_{2}} \right].
\end{equation}
\end{small}
By minimizing this objective, $\mathcal{G}_{\mathcal{T}}$ learns to generate quality-related features for specific distortion.

\textbf{Alignment loss.} During the training process, we synchronize $\mathcal{G}_{\mathcal{T}}$ with $\mathcal{G}_{\mathcal{T}-1}$, which means that the previous distortion features generated by $\mathcal{G}_{\mathcal{T}}$ should be the same as that generated by $\mathcal{G}_{\mathcal{T}-1}$. The generator alignment loss is defined as
\begin{small}
\begin{equation}
    L_{\mathcal{T}}^{GA}={{\mathbb{E}}_{\mathbf{z}\sim \mathcal{N}(\mathbf{0},\mathbf{I}), s \sim \bar{\mathcal{S}}, j \sim p_{j_{t<\mathcal{T}}} }}\left[ {{\left\| {{\mathcal{G}}_{\mathcal{T}}}(\mathbf{z},s,j)-{{\mathcal{G}}_{\mathcal{T}\text{-1}}}(\mathbf{z},s,j) \right\|}_{\text{2}}} \right].
\end{equation}
\end{small}
Similarly, we also apply alignment loss to the current discriminator, which encourages the quality prediction values and the real/fake probability towards previous distortion features to be the same as that of the former discriminator.
\begin{scriptsize}
\begin{equation}
\begin{aligned}
   L_{\mathcal{T}}^{DA}=&{{\mathbb{E}}_{\mathbf{z}\sim \mathcal{N}(\mathbf{0},\mathbf{I}), s \sim \bar{\mathcal{S}}, j \sim p_{j_{t<\mathcal{T}}} }}\left[ {{\left\| {{\mathcal{D}}_{\mathcal{T}}^{qua}}({{{\tilde{h}}}_{j}}^{\mathcal{T}-1},j)-{{\mathcal{D}}_{\mathcal{T}\text{-1}}^{qua}}({{{\tilde{h}}}_{j}}^{\mathcal{T}-1},j) \right\|}_{\text{2}}} \right]+\\&{{\mathbb{E}}_{\mathbf{z}\sim \mathcal{N}(\mathbf{0},\mathbf{I}), s \sim \bar{\mathcal{S}}, j \sim p_{j_{t<\mathcal{T}}} }}\left[ {{\left\| {{\mathcal{D}}_{\mathcal{T}}^{r/f}}({{{\tilde{h}}}_{j}}^{\mathcal{T}-1},j)-{{\mathcal{D}}_{\mathcal{T}\text{-1}}^{r/f}}({{{\tilde{h}}}_{j}}^{\mathcal{T}-1},j) \right\|}_{\text{2}}} \right],
\end{aligned}
\end{equation}
\end{scriptsize}
where ${{{\tilde{h}}}_{j}}^{\mathcal{T}-1}={{\mathcal{G}}_{\mathcal{T}-1}}(\mathbf{z},s,j)$ denotes the pseudo features of distortion $j$ generated by $\mathcal{G}_{\mathcal{T}-1}$.

\textbf{Full objective.}
When $\mathcal{T}>0$, the objective functions to optimize $G_{\mathcal{T}}$ and $D_{\mathcal{T}}$ are written respectively as:
\begin{equation}
L_{\mathcal{T}}^{\mathcal{G}}=-L_{\mathcal{T}}^{adv}+{{\lambda }_{qua}}L_{\mathcal{T}}^{qua\_r}+{{\lambda }_{align}}L_{\mathcal{T}}^{GA},
\label{equ:L_G}
\end{equation}
\begin{equation}
    L_{\mathcal{T}}^{\mathcal{D}}=L_{\mathcal{T}}^{adv}+{{\lambda }_{qua}}L_{\mathcal{T}}^{qua\_f}+{{\lambda }_{align}}L_{\mathcal{T}}^{DA},
\label{equ:L_D}
\end{equation}
where ${\lambda }_{qua}$ and ${\lambda }_{align}$ are hyper-parameters controlling the relative importance of quality prediction loss and alignment loss, compared to the adversarial loss. When $\mathcal{T}=0$, we don't have the alignment loss item.

\subsubsection{Training auxiliary multi-head regression network}
The feature extractor $\hat{U}_{\mathcal{T}}$ of the multi-head regression network $\hat{\mathcal{R}}_{\mathcal{T}}$ is initialized with the feature extractor  $U_{\mathcal{T}}$ of the single-head network $\mathcal{R}_{\mathcal{T}}$. The training objective is defined as:
\begin{equation}
    {{L}^{MSE2}}={{\mathbb{E}}_{(\mathbf{x},s)\sim {{D}^{\mathcal{T}}},j \sim p_{j_{t=\mathcal{T}}}}}\left[{{\left\| {{{\hat{\mathcal{R}}}}_{\mathcal{T}}}(\mathbf{x},j),s \right\|}_{2}}\right],
\end{equation}
where the feature extractor $\hat{U}_{\mathcal{T}}$ and the previous distortions' prediction heads $\hat{V}_{j_{t<\mathcal{T}}}$ of $\hat{\mathcal{R}}_{\mathcal{T}}$ are frozen and only the current task's distortion prediction heads $\hat{V}_{j_{t=\mathcal{T}}}$ are trainable.

\begin{table*}[htbp]
\centering
\caption{Description of IQA databases. DisNum refers to the number of synthetic distortion types. MOS refers to Mean Opinion Score and a higher value denotes better perceptual quality. DMOS refers to Differential Mean Opinion Score and is inversely proportional to MOS. DisImageNum refers to the number of the distorted images. }
\setlength{\tabcolsep}{1mm}{
\begin{tabular}{c|cccccc}
\hline
Database & Scenario & DisNum & Subjective Testing Methodology & Annotation & Range & DistImageNum  \\ \hline
LIVE\cite{LIVE} & Synthetic & 5 & Single stimulus continuous quality rating & DMOS & {[}0,100{]} & 779  \\
CSIQ\cite{CSIQ} & Synthetic & 6 & Multi stimulus absolute category rating & DMOS & {[}0,1{]} & 866  \\
KADID-10K\cite{KADID-10K} & Synthetic & 25 & Double stimulus absolute category rating with crowdsourcing & MOS & {[}1,5{]} & 10,125  \\ \hline
BID\cite{BID} & Authentic & - & Single stimulus continuous quality rating & MOS & {[}0,5{]} & 586 \\
CLIVE\cite{CLIVE} & Authentic & - & Single stimulus continuous quality rating with crowdsourcing & MOS & {[}0,100{]} & 1,162  \\
KonIQ-10K\cite{KonIQ-10k} & Authentic & - & Single stimulus absolute category rating with crowdsourcing & MOS & {[}1,5{]} & 10,073 \\ \hline
\end{tabular}}
\label{tab:dataset}
\end{table*}

\section{Experiments}
\label{sec:experiments}
\subsection{Datasets}
\label{sec:datasets}
As shown in Table \ref{tab:dataset}, we conduct experiments on six IQA datasets, among which three are synthetically distorted (LIVE \cite{LIVE}, CSIQ \cite{CSIQ}, KADID-10K \cite{KADID-10K}) and the others are authentically distorted (BID \cite{BID}, CLIVE \cite{CLIVE} and KonIQ-10K \cite{KADID-10K}).

The LIVE database includes 779 synthetically distorted images, which are generated from 29 reference images by corrupting them with five distortion types, i.e. JPEG-2000 compression, JPEG compression, white Gaussian noise, Gaussian blur, and fast fading rayleigh at five to eight intensity levels. DMOS of each distorted image ranges from 1 to 100 and is collected using the single stimulus continuous quality rating method. The CSIQ database consists of 866 synthetically distorted images which are derived from 30 original images distorted with six distortion types at four to five different intensity levels. The distortions are JPEG compression, JPEG-2000 compression, global contrast decrements, additive pink Gaussian noise, additive white Gaussian noise, and Gaussian blurring. The ratings are reported in the form of DMOS ranging from 0 to 1 using multi-stimulus absolute category rating method. KADID-10K consists of 10,125 distorted images derived from 81 pristine images considering 25 different distortion types at 5 intensity levels. The distortion types include Gaussian blur (GB), lens blur (LB), motion blur (MB), color diffusion (CD), color shift (CS), color quantization (CQ), two kinds of color saturation (CSA1 and CSA2), JPEG-2000 compression (JP2K), JPEG compression (JPEG), white noise (WN), white noise in color component (WNCC), impulse noise (IN), multiplicative noise (MN), denoise (DN),  brighten (BR), darken (DA), mean shift (MS), jitter (JIT), non-eccentricity patch (NEP), pixelate (PIX), quantization (Q), color block (CB), high sharpen (HS), and contrast change (CC). The MOS of each image ranges from 1 to 5 and is collected using double stimulus absolute category rating with crowdsourcing.

The BID database contains 586 authentically distorted pictures taken by human users in a variety of scenes, camera apertures, and exposition times. The distorted images are mostly blurred, which not only include typical, easy-to-model blurring cases but also more complex, realistic ones. The MOS of each image ranges from 0 to 5 and is collected using the single stimulus continuous rating method. The CLIVE database contains 1,162 authentically distorted images captured from diverse mobile devices. Each image is collected without artificially introducing any distortions beyond those occurring during capture, processing, and storage by a user's device. The MOS of each image ranges from 0 to 100 and is collected by the single stimulus continuous quality rating with crowdsourcing. The KonIQ-10K database consists of 10,073 authentically distorted images selected from a massive public multimedia database, YFCC100m \cite{YFCC100M}.  The MOS of each image ranges from 1 to 5 and is collected by the single stimulus absolute category rating with crowdsourcing.

\subsection{Implementation details}
From Section \ref{sec:datasets}, we can see that each dataset consists of various distortion types whose distributions vary a lot. Moreover, each dataset contains distinct distortion types and the quality scores is collected using different subjective testing methodology. Therefore, the shift of distortion types and datasets may both bring the risks of catastrophic forgetting.

In order to test the effectiveness of LIQA  facing with continuous distortion shift, we adopted KADID-10K dataset and randomly split the 25 distortion types into two groups, i.e. a base group and a novel group. The base group includes 7 distortion types (CQ, JPEG, CSA1, WNCC, Q, JIT and IN) and is used for training the base task. The novel group includes 18 distortion types (JP2K, CSA2, PIX, WN, CB, GB, DA, CC, BR, NEP, MS, MB, MN, LB, DN, HS, CD and CS) and can be divided into 18, 9 and 3  novel tasks with the incremental step $\Delta$ set to 1, 2 and 6 respectively in our experiments. Then we randomly permuted the distortions in the novel group to test the robustness to the distortion order.  Moreover, in order to test the effectiveness of LIQA when facing with continuous dataset shift, we sequentially add one dataset per task. We regarded each dataset as a kind of distortion and trained LIQA.

Following the work of UNIQUE \cite{zhang2021uncertainty}, we randomly sampled 80\% images from each dataset for training, 10\% for validation and the left 10\% for testing. Specially, for the three synthetic datasets, we split the datasets according to the reference images in order to ensure content dependence. During training, we randomly cropped the images into $300\times300$ and during validation and testing, we cropped the images to $300\times300$ in the center. For each task, we trained the single-head regression network for 70 epochs, the generator and the discriminator for 500 epochs and the multi-head regression network for 70 epochs. We adopted an early-stopping strategy and chose the model that performed the best on the validation set for testing. Specially, we selected the best-performing network after 15 epochs to make sure the learning of the current task. For training the generator, we adopted a data augmentation strategy and expand the size of the original dataset tenfold offline, because the image for each distortion type are too few to train a good generator.  But for training the regression network, we adopted the original dataset. Each experiment was run five times and the results were averaged.

 $\lambda_{FD}$,  $\lambda_{PR}$ and $\lambda_{MSE}$ in equation \ref{equ: full_single} are  emperically set to 0.001, 10.0, 1.0 respectively. We set
${\lambda }_{qua}$ and ${\lambda }_{align}$ in equations \ref{equ:L_G} and \ref{equ:L_D} to 1.0 and 3.0.
The learning rate of the feature extractor and the prediction heads for regression networks was set to $1e^{-4}$ when learning the base task. We lowered down the learning rate of the feature extractor to $1e^{-6}$ when learning the novel tasks. We linearly re-scaled the subjective scores of each of the six databases to [0,1], where higher value denotes better perceptual quality. The regression networks were trained using Adam with a batch size of 48 and the buffer size of pseudo features per batch was set to 1400. To generate pseudo features for each previous distortion, we split the re-scaled quality range into five interval segments, i.e. [0,0.2), [0.2,0.4), [0.4,0.6), [0.6,0.8), [0.8,1.0]. For each quality interval, we allocated $1400/M_{pre}/5$ pseudo features to make sure that the generated pseudo features can well cover the distribution of the real features. The generator as well as the discriminator were trained using Adam with a batch size of 128.

\subsection{Compared methods}

We compare LIQA with fine-tuning, joint training and three prior-focused  regularization-based lifelong learning approaches, i.e. EWC \cite{ewc}, online EWC \cite{onlineewc} and SI \cite{si}.  All the compared approaches were trained with Adam with a batch size of 48 and the learning rate of $1e^{-4}$. We chose the best-performed model during validation for testing just like the training process of LIQA.

\textbf{Fine-tuning (FT)}: Modifying the parameters of an existing network to adapt to a new task. At current task $\mathcal{T}_{\mathcal{T}}$, we directly fine-tune the single-head regression network initialized by $\mathcal{R}_{\mathcal{T}-1}$ using the current task's training data $D_{\mathcal{T}}$. The loss of FT is defined as:
\begin{equation}
     L_{\mathcal{T}}^{FT}={{\mathbb{E}}_{(\mathbf{x},s)\sim {{D}_{\mathcal{T}}}}}\left[ {{\left\| {{\mathcal{R}}_{\mathcal{T}}}(\mathbf{x})-s \right\|}_{2}} \right].
\end{equation}

\textbf{EWC}: EWC estimates the importance of parameter $i$ by the $i$-th diagonal element of the current task $T_{\mathcal{T}}$'s Fisher Information matrix, the regularization loss is defined as
\begin{equation}
    {{L}^{reg-{EWC}}}=\frac{1}{2}\sum\limits_{t=0}^{\mathcal{T}-1}{(}\sum\limits_{i=1}^{{{N}_{params}}}{\mathbf{F}_{ii}^{t}{{({{\theta }_{i}}-\hat{\theta }_{i}^{t})}^{2}}}),
\end{equation}
where $\hat{\theta }_{i}^{t}$ is the value of the parameter $i$ after finishing training on task $T_{t}$. $\mathbf{F}_{ii}$ can be calculated by
\begin{equation}
    \mathbf{F}_{ii}^{t}=\frac{1}{\left| {{D}^{t}} \right|}\sum\limits_{(\mathbf{x},s)\sim {{D}^{t}}}{(\frac{\partial {{L}_{2}}({{\mathcal{R}}_{t}}(\mathbf{x};{{{\hat{\theta }}}^{t}}),s)}{\partial {{\theta }_{i}}}}{{)}^{2}},
\end{equation}
where $L_{2}$ denotes $L_{2}$ loss function. The full objective of EWC is:
\begin{equation}
    L_{\mathcal{T}}^{EWC}=L_{\mathcal{T}}^{FT}+\lambda_{EWC}{{L}^{reg-{EWC}}}.
\end{equation}
In our experiment, we emperically set $\lambda_{EWC}$ to 5000.0 via a coarse grid search on the held-out validation set following the work of \cite{si}.

\textbf{Online EWC}: The regularization term for online EWC is given by:
\begin{equation}
    L_{\mathcal{T}}^{reg-onlineEWC}=\sum\limits_{i=1}^{{{N}_{params}}}{\mathbf{\tilde{F}}_{ii}^{\mathcal{T}-1}{{({{\theta }_{i}}-\hat{\theta }_{i}^{(\mathcal{T}-1)})}^{2}}},
\end{equation}
where $\mathbf{\tilde{F}}_{ii}^{\mathcal{T}-1}$ is a running sum of the $i$-th diagonal elements of the Fisher Information matrices of tasks ${{T}_{t<={\mathcal{T}-1}}}$, i.e. $\mathbf{\tilde{F}}_{ii}^{\mathcal{T}-1}=\gamma \mathbf{\tilde{F}}_{ii}^{\mathcal{T}-2}$ + $\mathbf{F}_{ii}^{\mathcal{T}-1}$. $\gamma <=1$ is a hyperparameter that governs the gradual decay of the contributions of previous tasks. The full objective of online EWC is given by:
\begin{equation}
    L_{\mathcal{T}}^{onlineEWC}=L_{\mathcal{T}}^{FT}+\lambda_{onlineEWC}{{L}^{reg-{onlineEWC}}}.
\end{equation}
In our experiment, we emperically set $\gamma$ to 1 and $\lambda_{onlineEWC}$ to 5000.0.

\textbf{SI}: SI estimates the importance for each parameter and protect the parameters important to previous tasks from changing. SI loss is defined as:
\begin{equation}
    \label{equ:si}
    L_{\mathcal{T}}^{reg-SI}=\sum\limits_{i=1}^{{{N}_{params}}}{\Omega _{i}^{(\mathcal{T}-1)}{{({{\theta }_{i}}-\hat{\theta }_{i}^{(\mathcal{T}-1)})}^{2}}},
\end{equation}
where $\hat{\theta }_{i}^{(\mathcal{T}-1)}$ is the value of parameter $i$ after finishing training on task $T_{\mathcal{T}-1}$. $\Omega _{i}^{(\mathcal{T}-1)}$ is the estimated importance of parameter $i$ for all the previous tasks:
\begin{equation}
    \Omega _{i}^{(\mathcal{T}-1)}=\sum\limits_{t=0}^{\mathcal{T}-1}{\frac{\omega _{i}^{t}}{{{(\Delta _{i}^{t})}^{2}}+\xi }},
\end{equation}
where $\Delta _{i}^{t}={{\hat{\theta }}_{i}}[{{N}_{iters}}^{t}]-{{\hat{\theta }}_{i}}[{{0}^{t}}]$, ${{N}_{iters}}^{t}$ is the number of iterations and ${{\hat{\theta }}_{i}}[{{0}^{t}}]$ indicates the value of parameter $i$ right before starting training on task $T_{t}$.  $\xi$ is a small value (usually set to 0.1). $\omega_{i}^{t}$ counts the contribution of parameter $i$ to the change in loss:
\begin{equation}
    \omega _{i}^{t}=\sum\limits_{n=1}^{{{N}_{iters}}}{({{{\hat{\theta }}}_{i}}[{{n}^{t}}]-{{{\hat{\theta }}}_{i}}[{{(n-1)}^{t}}])\frac{-\partial {{L}_{total}}[{{n}^{t}}]}{\partial {{\theta }_{i}}}},
\end{equation}
where ${{{\hat{\theta }}}_{i}}[{{n}^{t}}]$ denotes the value of the parameter $i$ after the $n$-th training iteration.

The full objective of SI is given by:
\begin{equation}
    L_{\mathcal{T}}^{SI}=L_{\mathcal{T}}^{FT}+\lambda_{SI}{{L}^{reg-{SI}}}.
\end{equation}
We emperically set $\lambda_{SI}$ to 100.0 in our experiments via a coarse grid search on the held-out validation set following the work of \cite{si}.

\begin{figure*}[htbp]
 \centering
 \subfigure[]{
  \label{fig:7base_1stepC_order1}
  \includegraphics[width = .49\textwidth]{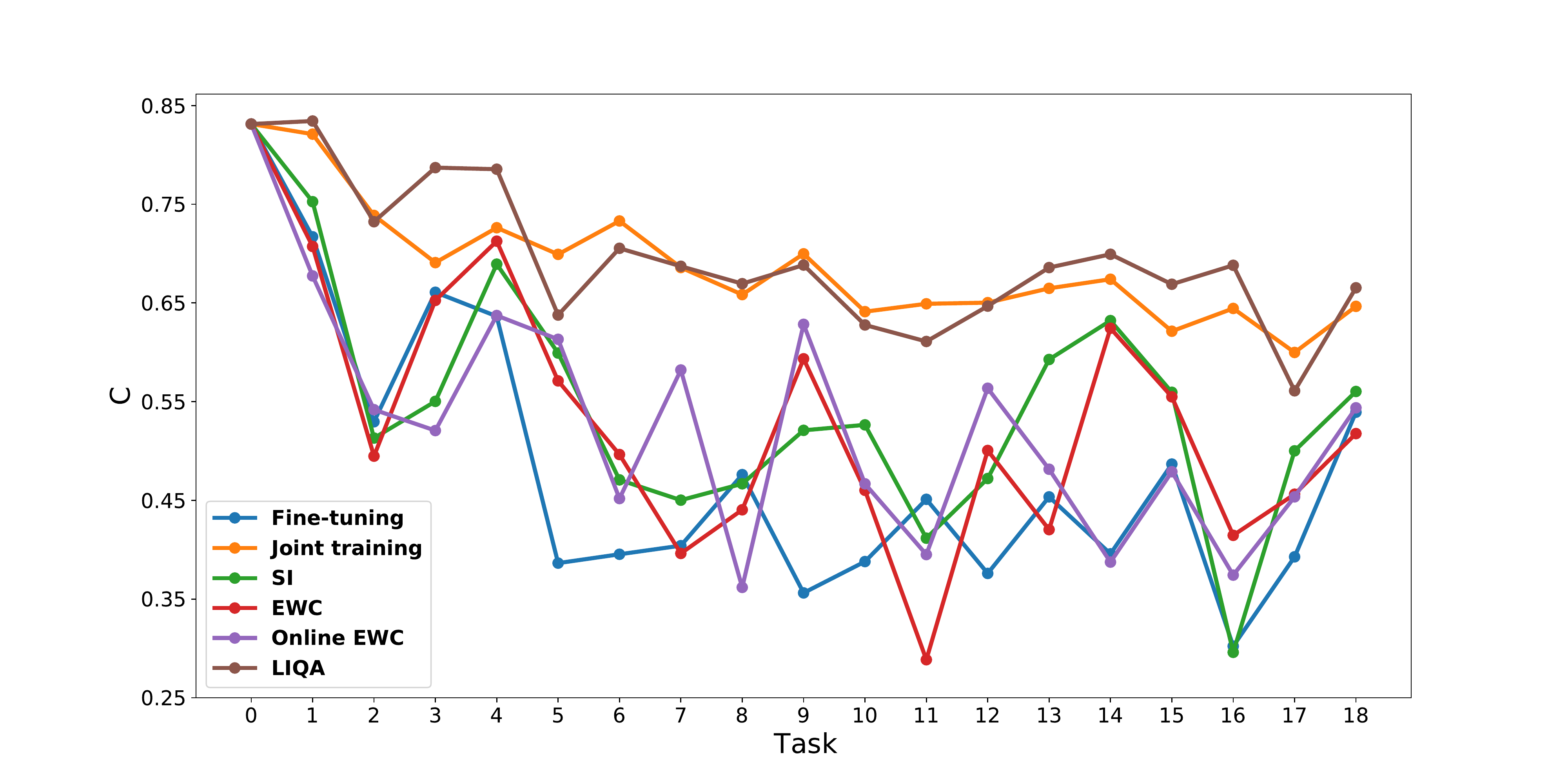}}
 \subfigure[]{
  \label{fig:7base_1stepF_order1}
  \includegraphics[width = .49\textwidth]{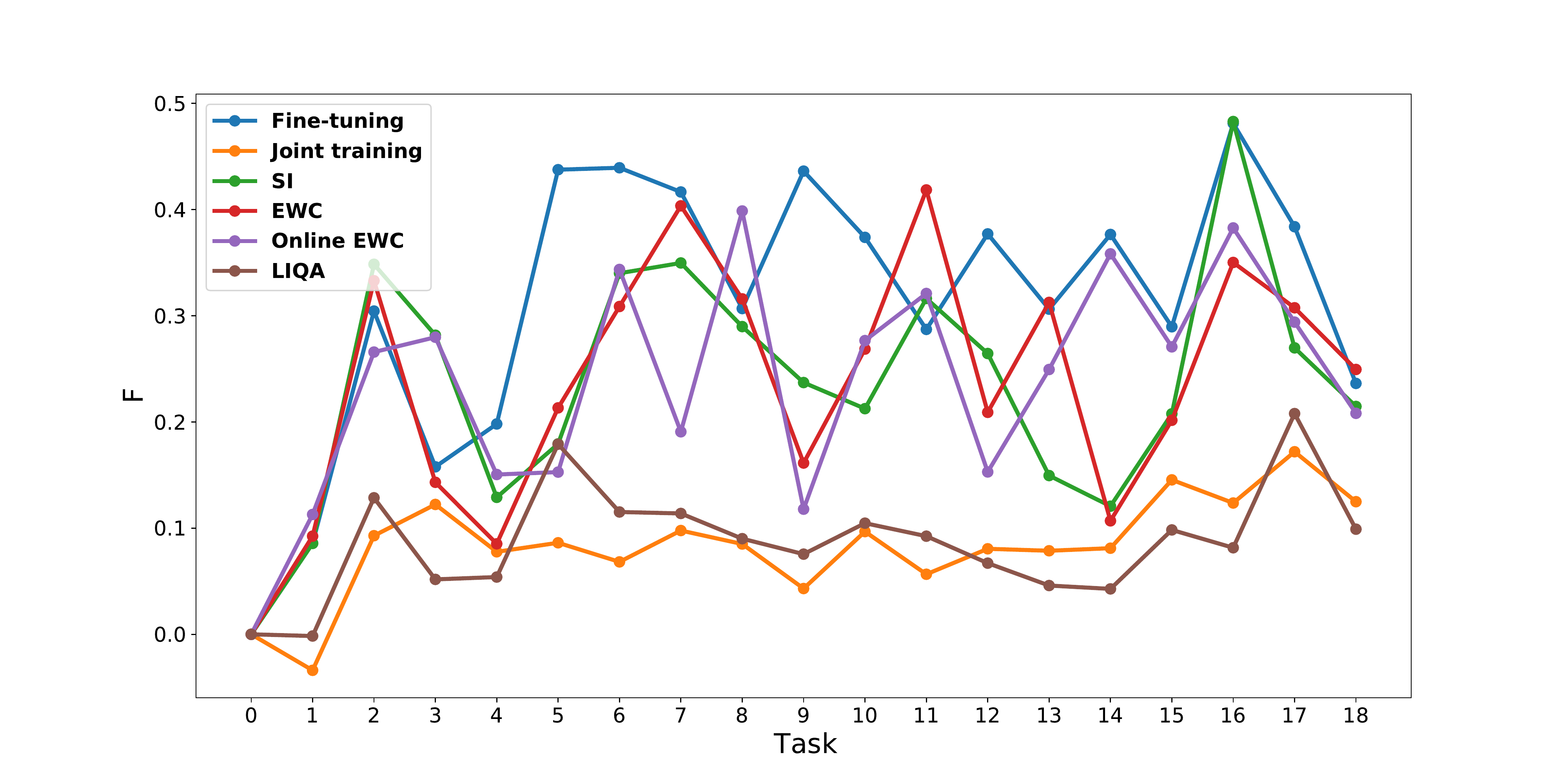}}

\caption{Performance of distortion shift with incremental step set to 1. (a) Correlation index with respect to tasks. (b) Forgetting index with respect to tasks.  }
\end{figure*}

\textbf{Joint training (JT)}: All the previously learned tasks' training data is stored and combined with the current task's training data  for training $\mathcal{R}_{\mathcal{T}}$. $\mathcal{R}_{\mathcal{T}}$  is initialized with $\mathcal{R}_{\mathcal{T}-1}$. The loss of JT is defined as:
\begin{equation}
     L_{\mathcal{T}}^{JT}={{\mathbb{E}}_{(\mathbf{x},s)\sim {{D}_{t<=\mathcal{T}}}}}\left[ {{\left\| {{\mathcal{R}}_{\mathcal{T}}}(\mathbf{x})-s \right\|}_{2}} \right].
\end{equation}
It should be noted that JT can be regarded as an upper bound of lifelong learning methods which are not allowed to get access to previous training data $D_{t<\mathcal{T}}$.
\begin{table*}[]
\centering
\caption{Performance comparison across various distortion types of KADID-10K at the final task. The best performance of each distortion among fine-tuning and the lifelong learning methods is highlighted in bold. }
\label{tab:last_session_distshift}
\begin{tabular}{c|ccccccccccccc}
\hline
 & CQ & JPEG & CSA1 & WNCC & Q & JIT & IN & JP2K & CSA2 & PIX & WN & CB & GB \\ \hline
FT & 0.450 & 0.530 & 0.077 & 0.898 & 0.678 & 0.746 & 0.736 & 0.270 & 0.802 & 0.075 & \textbf{0.838} & 0.300 & 0.708 \\
EWC & 0.435 & 0.711 & 0.065 & 0.782 & 0.672 & 0.635 & 0.681 & 0.590 & 0.653 & 0.494 & 0.693 & \textbf{0.380} & 0.470 \\
Online EWC & 0.471 & 0.782 & 0.164 & 0.768 & 0.697 & 0.771 & 0.563 & 0.679 & 0.701 & 0.466 & 0.575 & 0.282 & 0.756 \\
SI & 0.235 & 0.794 & 0.060 & 0.710 & 0.658 & 0.789 & 0.454 & 0.741 & 0.654 & 0.433 & 0.629 & 0.318 & \textbf{0.870} \\
LIQA & \textbf{0.622} & \textbf{0.825} & \textbf{0.403} & \textbf{0.903} & \textbf{0.827} & \textbf{0.864} & \textbf{0.879} & \textbf{0.855} & \textbf{0.791} & \textbf{0.627} & 0.825 & 0.341 & 0.869 \\ \hline
\rowcolor[HTML]{EFEFEF}
JT & 0.515 & 0.814 & 0.378 & 0.825 & 0.398 & 0.886 & 0.864 & 0.851 & 0.731 & 0.680 & 0.796 & 0.221 & 0.900 \\ \hline
 & DA & CC & BR & NEP & MS & MB & MN & LB & DN & HS & CD & CS & Avg \\ \hline
FT & \textbf{0.546} & \textbf{0.136} & 0.278 & 0.122 & \textbf{0.402} & 0.385 & 0.858 & \textbf{0.814} & 0.867 & 0.584 & 0.483 & 0.899 & 0.539 \\
EWC & 0.393 & 0.116 & \textbf{0.597} & 0.128 & 0.362 & 0.230 & 0.772 & 0.361 & 0.829 & 0.614 & 0.390 & 0.887 & 0.518 \\
Online EWC & 0.525 & 0.035 & 0.237 & \textbf{0.236} & 0.325 & 0.455 & 0.773 & 0.620 & 0.806 & 0.503 & 0.573 & 0.830 & 0.544 \\
SI & 0.516 & 0.115 & 0.458 & 0.162 & 0.366 & \textbf{0.618} & 0.770 & 0.764 & 0.797 & 0.561 & \textbf{0.589} & \textbf{0.950} & 0.560 \\
LIQA & 0.522 & 0.047 & 0.549 & 0.214 & 0.373 & 0.579 & \textbf{0.917} & 0.786 & \textbf{0.949} & \textbf{0.783} & 0.519 & 0.766 & \textbf{0.665} \\ \hline
\rowcolor[HTML]{EFEFEF}
JT & 0.357 & 0.123 & 0.711 & 0.283 & 0.047 & 0.900 & 0.807 & 0.854 & 0.858 & 0.844 & 0.728 & 0.797 & 0.647 \\ \hline
\end{tabular}

\end{table*}

\subsection{Performance with respect to distortion shift}
\label{subsec:distortion_shift}
We first evaluate the performance of LIQA when faced with continuous distortion shift on KADID-10K.   We respectively set the incremental step $\Delta$ to 1, 2 and 6 and quantify the performance by computing  the correlation index $C$ and the forgetting index $F$ of each task. The network is first trained on the base task which consists of 7 base distortion types and then sequentially trained on novel tasks.

When the incremental step is set to 1,  we sequentially add 18 novel distortions following the  permutation order of JP2K$\to$CSA2$\to$PIX$\to$WN$\to$CB$\to$GB$\to$DA$\to$CC$\to$BR$\to$ NEP$\to$ MS$\to$ MB$\to$ MN$\to$ LB$\to$DN$\to$HS$\to$ CD$\to$ CS, leading to 18 novel tasks.  The correlation index and the forgetting index with respect to tasks  are shown in Fig. \ref{fig:7base_1stepC_order1} and Fig. \ref{fig:7base_1stepF_order1} respectively, from which we can see that both the correlation index and the forgetting index of fine-tuning are very unstable and the performances of previous tasks are easily influenced by the current task. Actually, setting the incremental step  to 1 is hard for the network to continually learn new knowledge. It is because that the number of per distortion's training data is limited (only 325 images).  Fine-tuning the network on such limited training data will cause the over-fitting problem and make the network easily be biased towards the current distortion. Intuitively, the network generalization ability will be impaired and the performances of the the previously seen distortions will be influenced. When the current distortion distribution varies greatly from the previous distortions, the performance of previous distortions will drop drastically. When the current distortion distribution resembles some previous distortions', the performance of previous distortion may not undergo drastic changes. Taking the task\#5 for example, when sequentially adding the training data of  CB, the performance of the most distortions will be seriously impaired for fine-tuning. The correlation index of task\#5 drops from 0.636 to 0.386 and the forgetting index increases from 0.198 to 0.437 compared with task\#4.  It is because that CB is a local distortion, whose distortion distribution varies greatly from that of the previously seen global distortions. Directly fine-tuning the network parameters on the training data of CB will make the network biased towards the currently learning distortion, leading to catastrophic forgetting of other distortion types. In contrast with fine-tuning, lifelong learning methods apparently mitigate the catastrophic forgetting phenomenon at task\#5. For EWC, online EWC, SI and LIQA, the forgetting index is reduced to  0.213, 0.153, 0.179 and 0.179 respectively.

Taking a look at the whole incremental learning process, we can find that the performance of  LIQA keeps stable while the performances of  other lifelong learning methods will change drastically at certain task session. Taking  task\#11 (adding MS) and task\#16 (adding HS) for example, the correlation index of EWC, online EWC and SI even cannot catch up with that of fine-tuning. In contrast, LIQA which takes advantages of replaying pseudo features to consolidate the learned knowledge of previously learned distortions can well resist the negative effect of certain task and obtain better robustness and stability when facing with continuous distortion shift. Without access to the previous training data, LIQA can obtain comparable performance with joint training.  Moreover, one thing that should be noted is that even if joint training utilizes training data of all seen task, adding new distortion type will also brings slight forgetting of previous distortion types, due to the conflict between different distortion types.

The SRCCs of each distortion type at the final task with incremental step set to 1 are listed in Table \ref{tab:last_session_distshift}, from which we can see that the performances of most distortions of LIQA outperform that of the other lifelong learning methods. By comparing  LIQA with joint training, we can find that the performances of some distortions (i.e. CQ, JPEG, CSA1, WNCC, Q, IN, JP2K, CSA2,  WN, CB, DA, MS, MN, DN) are better than that of joint training and the average SRCC of all seen distortions is also better than that of joint training. The performance of the final task  illustrates that LIQA has better ability to preserve the previously learned knowledge (the performances of the 7 base distortions of LIQA are comparable with that of joint training). Besides, consolidating the learned knowledge does not interfere with the learning of new knowledge (the performances of the novel distortions are satisfactory).

\begin{figure*}[htbp]
 \centering
\subfigure[]{
  \label{fig:7base_2stepC_order1}
  \includegraphics[width = .49\textwidth]{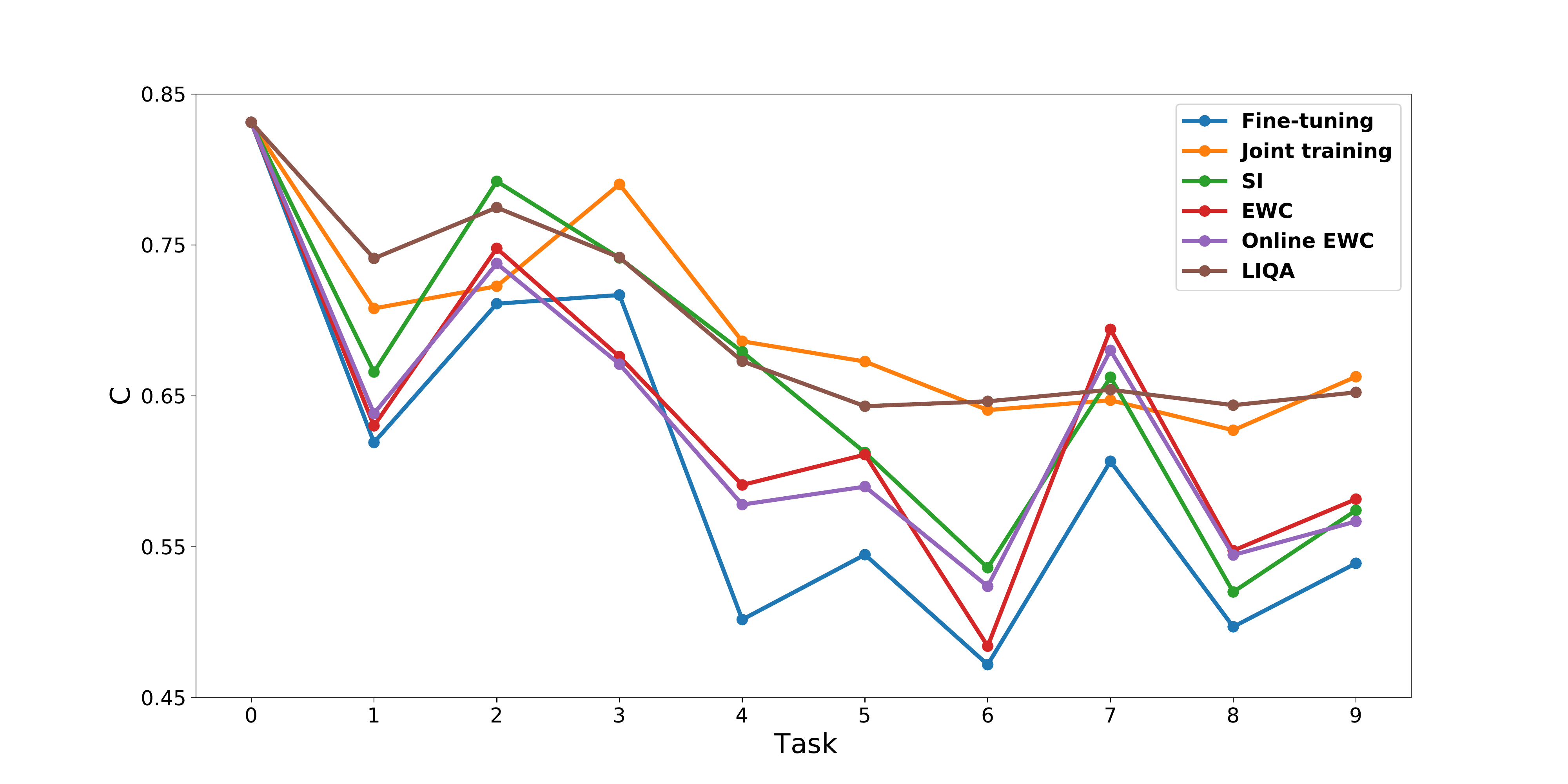}}
 \subfigure[]{
  \label{fig:7base_2stepF_order1}
  \includegraphics[width = .49\textwidth]{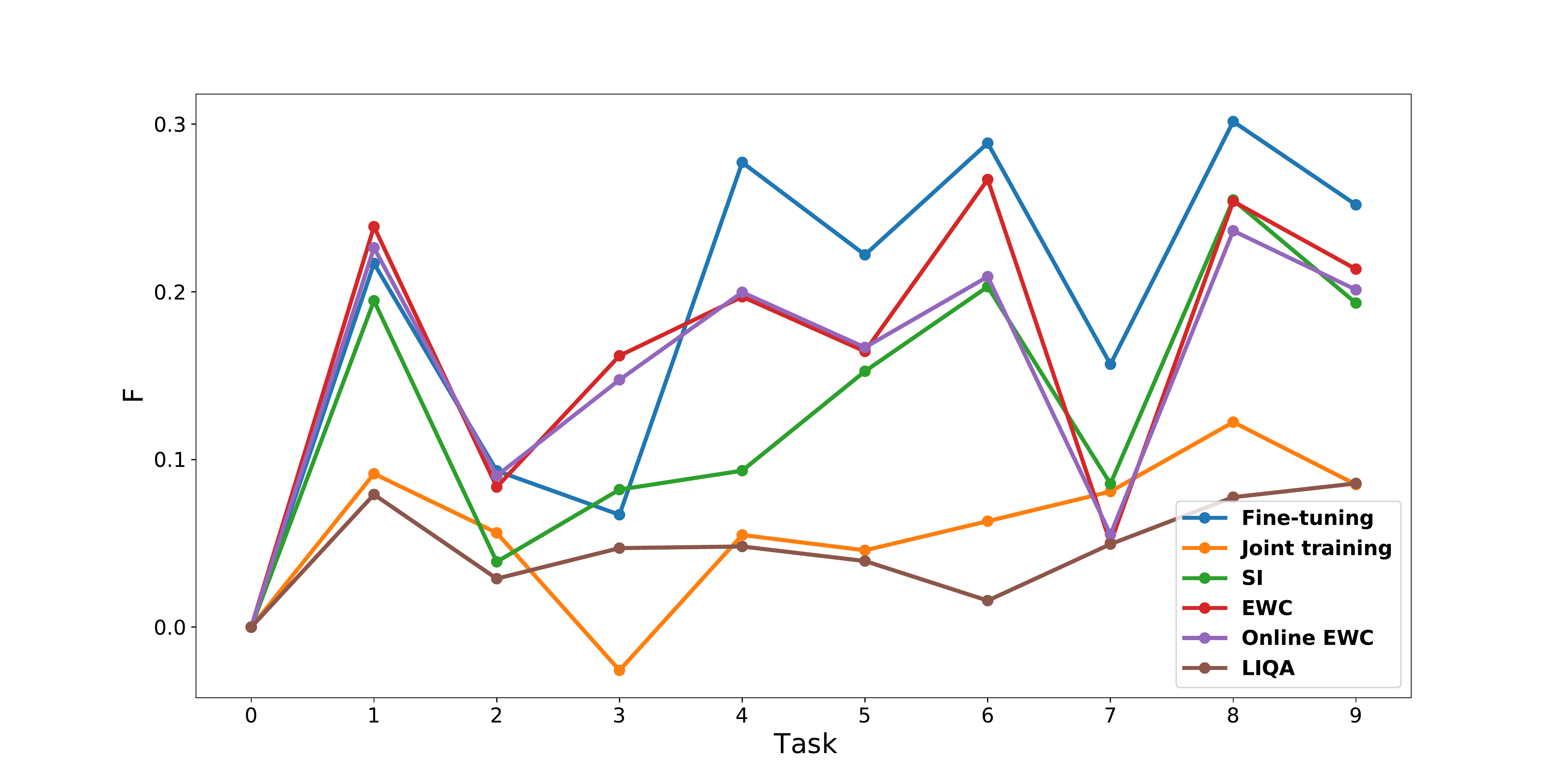}}
\caption{Performance of distortion shift with incremental step set to 2. (a) Correlation index with respect to task session. (b) Forgetting index with respect to task session.  }
\end{figure*}

When the incremental step is set to 2, we add two distortions per task following the order of (JP2K, CSA2)$\to$(PIX, WN)$\to$(CB, GB)$\to$(DA, CC)$\to$(BR, NEP)$\to$(MS, MB)$\to$ (MN,LB)$\to$(DN, HS)$\to$(CD, CS), leading to 9 novel tasks. The correlation index and the forgetting index with respect to tasks are shown in Fig. \ref{fig:7base_2stepC_order1} and Fig. \ref{fig:7base_2stepF_order1} respectively. By comparing Fig. \ref{fig:7base_2stepC_order1} and Fig. \ref{fig:7base_1stepC_order1}, we can find that when each task contains 2 distortions, the catastrophic forgetting of fine-tuning becomes less serious. The worst correlation index of fine-tuning in Fig. \ref{fig:7base_2stepC_order1} is 0.47 while the worst correlation index of fine-tuning in Fig. \ref{fig:7base_1stepC_order1} is 0.30. The biggest forgetting index in Fig. \ref{fig:7base_2stepF_order1} of fine-tuning  decreases from 0.48 to 0.30 compared with Fig. \ref{fig:7base_1stepF_order1}.  It is because that the negative effect of certain distortion will be weaken by another distortion. Taking the task\#3 in Fig. \ref{fig:7base_2stepC_order1} for example,  combination of CB and GB  will not seriously impair the performance of previous distortions.
By comparing the forgetting index of all methods shown in Fig. \ref{fig:7base_2stepF_order1}, we can find that the forgetting of LIQA is low and stable, and can well preserve the learned knowledge while learning new tasks. In contrast, the forgetting indexes of EWC, online EWC and SI are unstable and sometimes very high (e.g. task\#1, task\#6, and task\#8).

\begin{figure*}[htbp]
 \centering
  \subfigure[]{
  \label{fig:7base_6stepC_order1}
  \includegraphics[width = .49\textwidth]{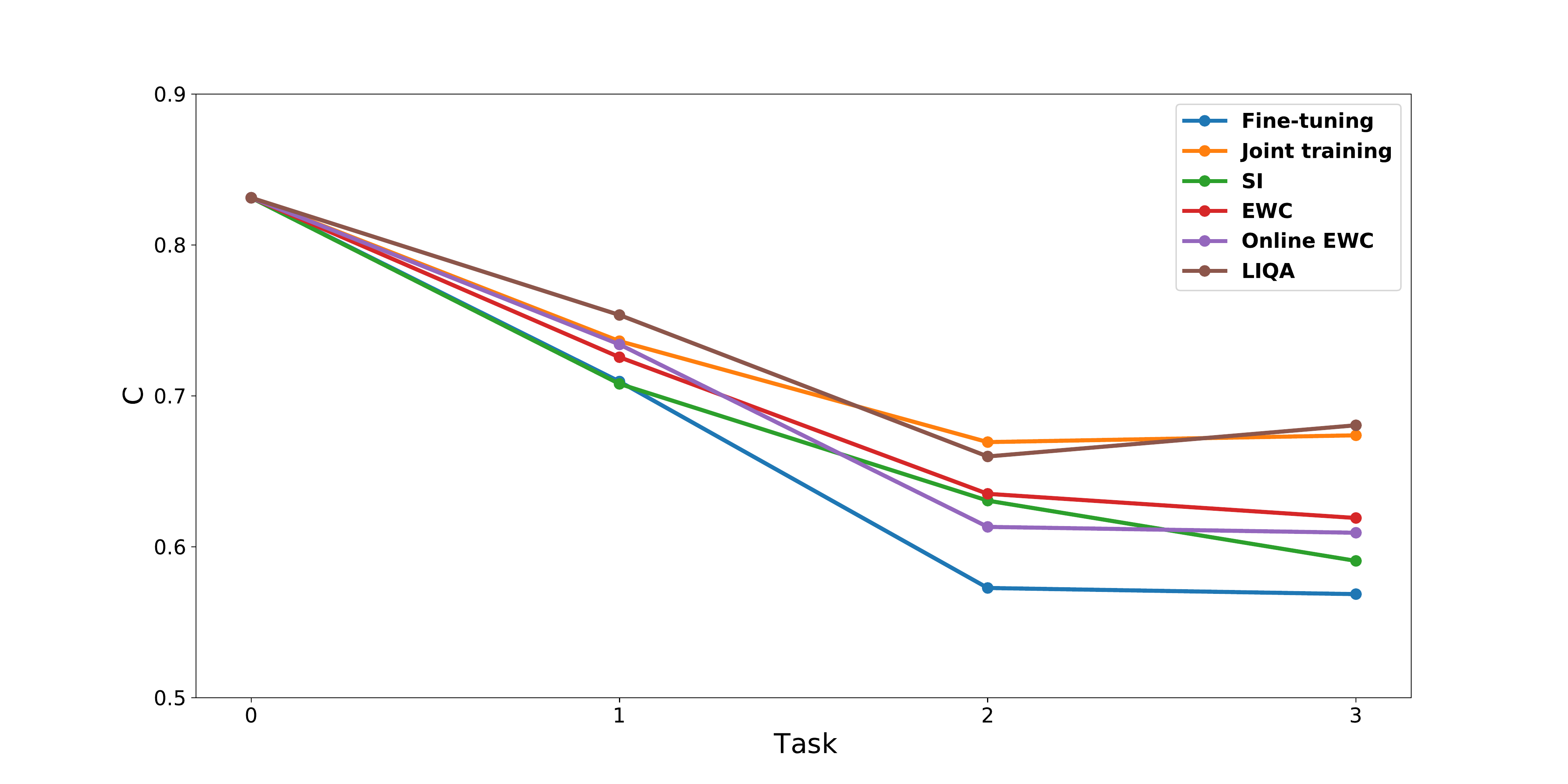}}
 \subfigure[]{
  \label{fig:7base_6stepF_order1}
  \includegraphics[width = .49\textwidth]{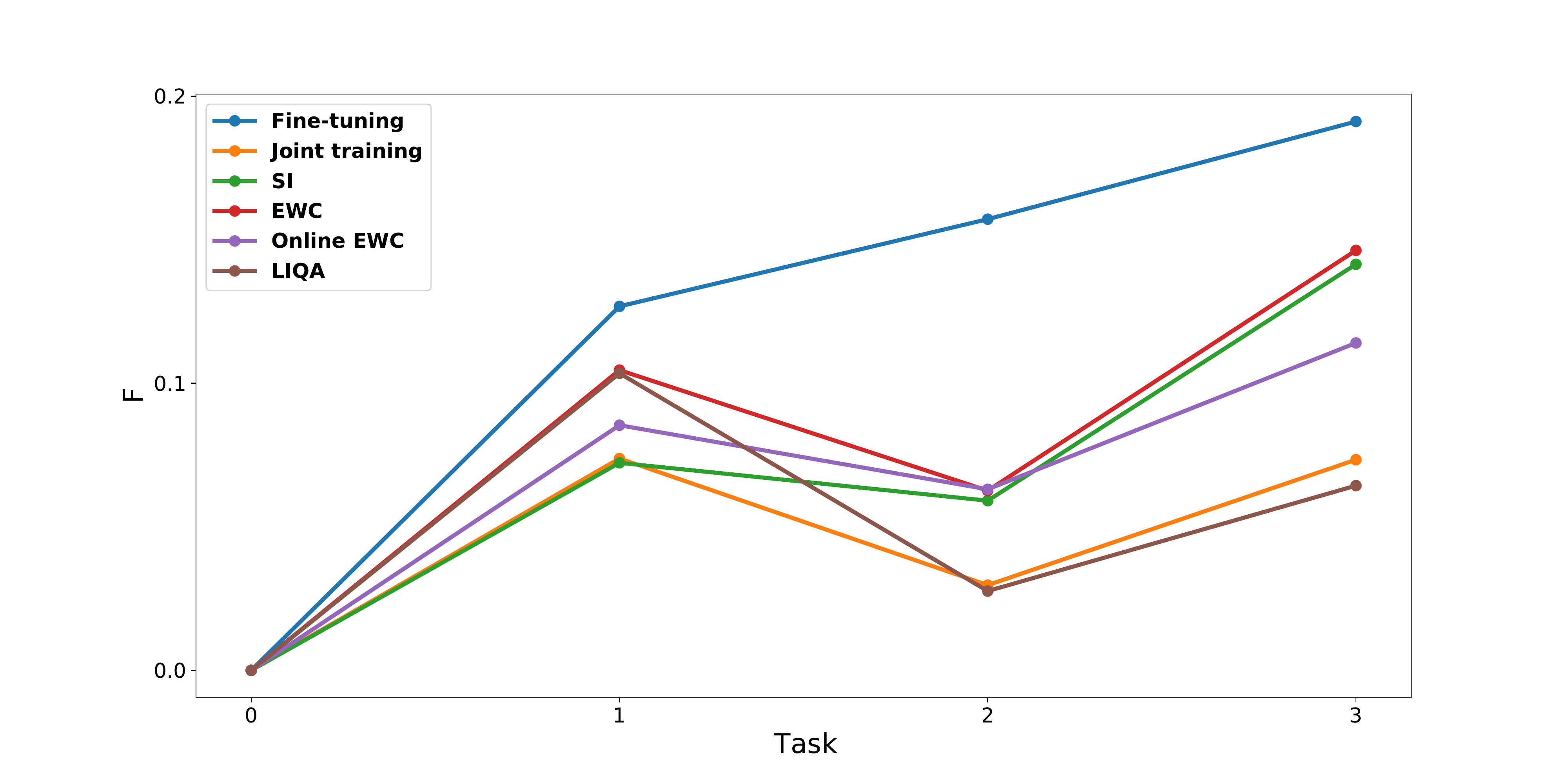}}
\caption{Performance of distortion shift with incremental step set to 6. (a) Correlation index with respect to tasks. (b) Forgetting index with respect to tasks.  }
\end{figure*}

When the incremental step is set to 6, we add two distortions per task following the order of (JP2K, CSA2, PIX, WN, CB, GB)$\to$(DA, CC, BR, NEP, MS, MB)$\to$ (MN,LB, DN, HS, CD, CS), leading to 3 novel tasks. The correlation index and the forgetting index with respect to tasks are shown in Fig. \ref{fig:7base_6stepC_order1} and Fig. \ref{fig:7base_6stepF_order1} respectively. From Fig. \ref{fig:7base_6stepC_order1} we can observe that the performance gap between fine-tuning and lifelong learning methods further shrinks. By comparing Fig. \ref{fig:7base_2stepF_order1} and  Fig. \ref{fig:7base_6stepC_order1}, we can find  that the biggest forgetting index of fine-tuning at certain task session further decreases from 0.30 to 0.19. It is because that as the distortion types and the number of the training data increases, the network generalization ability is also improved. The catastrophic forgetting caused by  certain distortion will be further suppressed.

To sum up, when the incremental step is set to 1, catastrophic problem will easily emerge. It is because the limited training data of certain distortion will make the network parameters over-fitted to the current task and destroy the generalization ability of the network. As the incremental step increases, the catastrophic forgetting of fine-tuning will be mitigated because the generalization ability is also improved. EWC, online EWC and SI can mitigate the catastrophic forgetting to some extent but the performance is unstable. LIQA utilizes the pseudo features to consolidate the learned knowledge, and can strike a good balance between the stability and the plasticity.

\subsection{Performance with respect to dataset shift}
\begin{figure*}[htbp]
 \centering

 \subfigure[]{
  \label{fig:crossdatasetC_order1}
  \includegraphics[width = .49\textwidth]{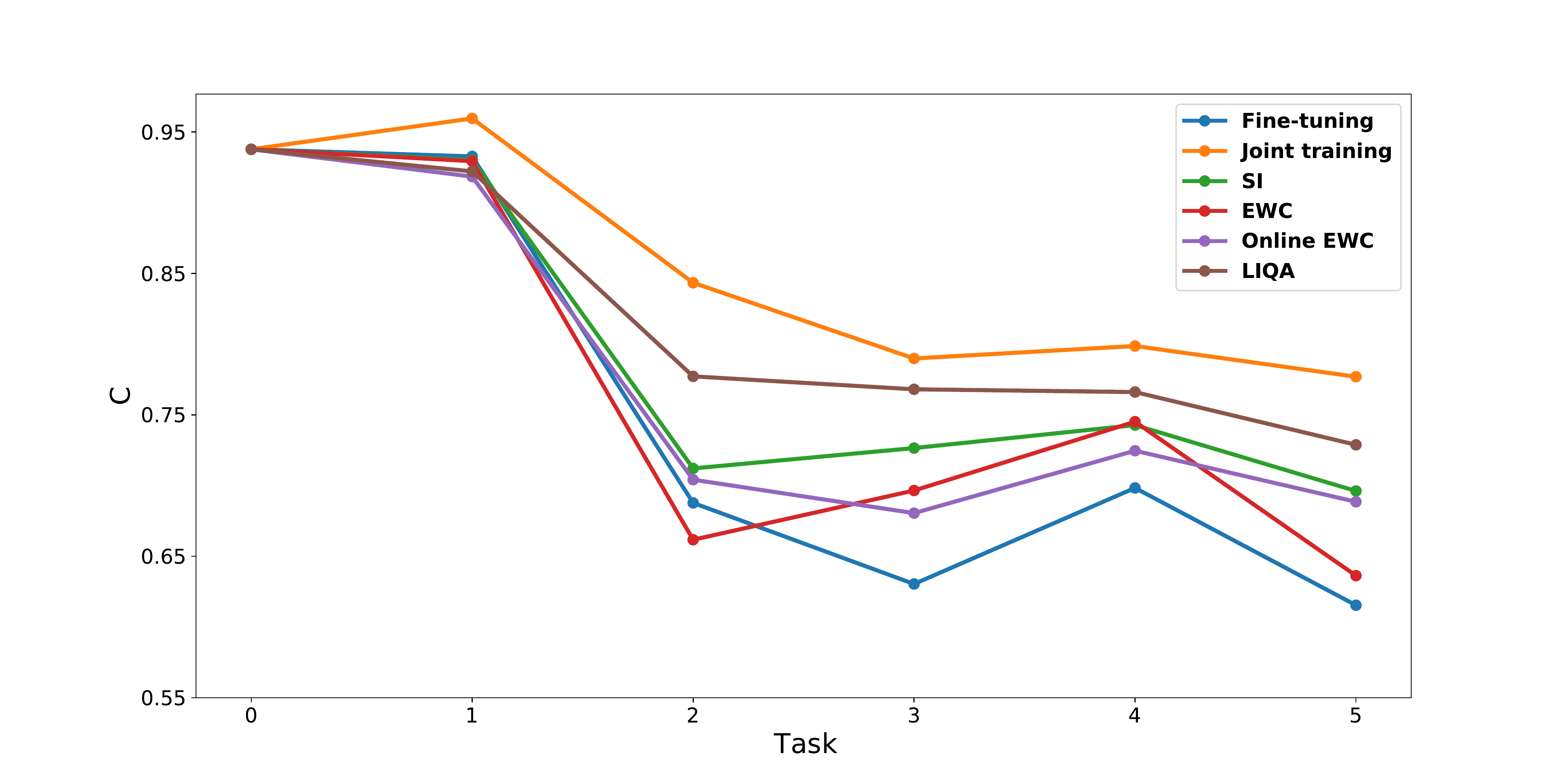}}
 \subfigure[]{
  \label{fig:crossdatasetF_order1}
  \includegraphics[width = .49\textwidth]{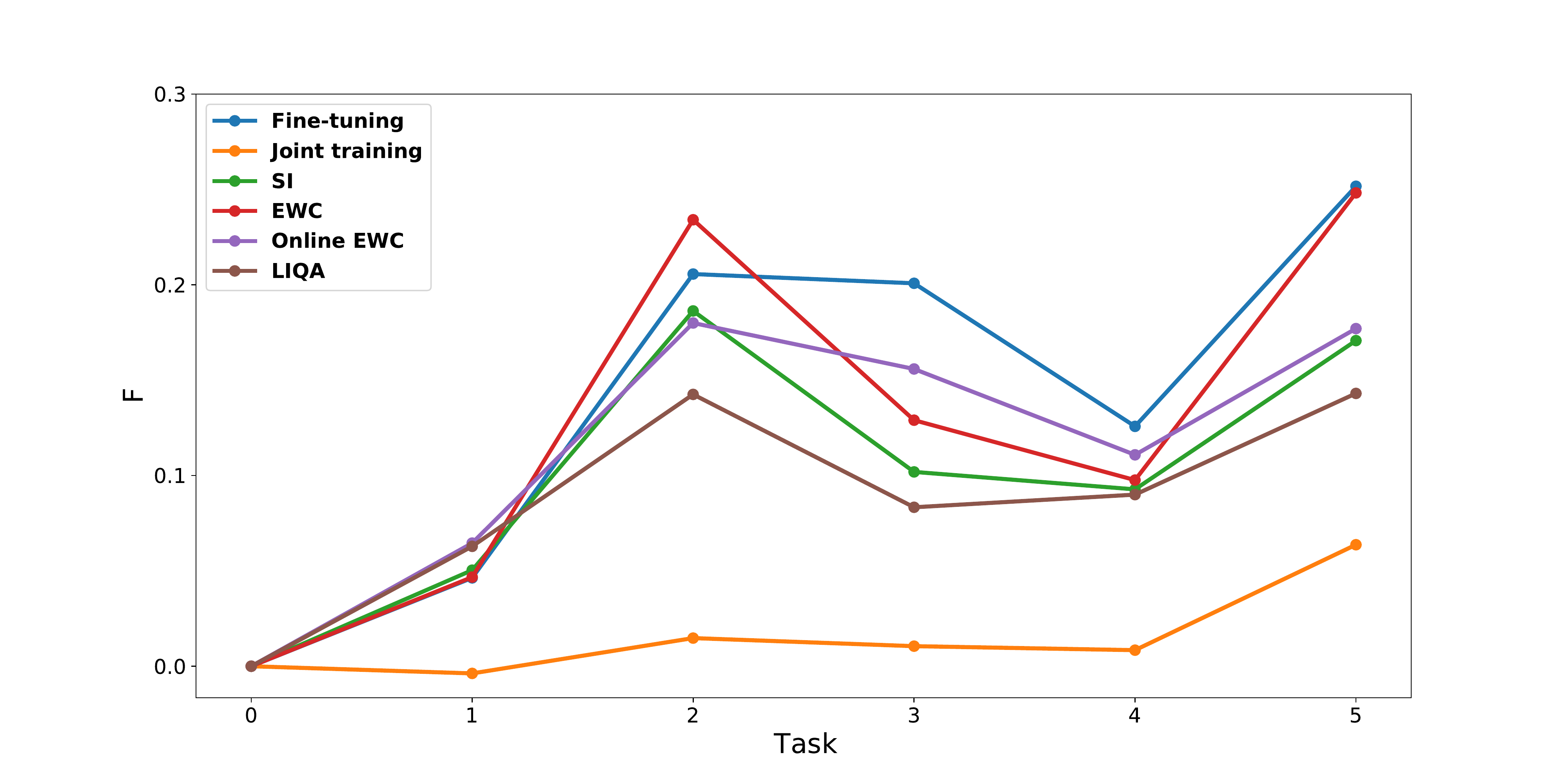}}

\caption{Performance of dataset shift with incremental step set to 1. (a) Correlation index with respect to tasks. (b) Forgetting index with respect to tasks. }
\end{figure*}

To evaluate the performance of LIQA when faced with continuous dataset shift, we conduct experiments on six IQA datasets following the permutation order of LIVE$\to$CSIQ$\to$BID$\to$CLIVE$\to$KonIQ-10K$\to$KADID-10K. Similar to the experiments with respect to distortion shift, we regard one dataset as one distortion type and set the incremental step to 1. The correlation index and the forgetting index are shown in Fig. \ref{fig:crossdatasetC_order1} and Fig. \ref{fig:crossdatasetF_order1} respectively. From Fig. \ref{fig:crossdatasetC_order1} we can see that the shift from LIVE to CSIQ  does not bring apparent forgetting for fine-tuning. It is because that LIVE and  CSIQ both include synthetically distorted images and have 4 overlapped distortion types (JPEG-2000 compression, JPEG compression, white Gaussian noise and Gaussian blur). However, when adding BID dataset which includes authentically distorted images, the forgetting indexes of fine-tuning and all the lifelong learning methods become apparent. It is known that the data distribution of BID dataset varies greatly from that of LIVE and CSIQ, and the apparent change makes the network biased towards the currently learning dataset, thus imparing the generalization ability towards other datasets. Similarly, we can find from Fig. \ref{fig:crossdatasetF_order1} that the forgetting index increases during the shift from authentic dataset to synthetic dataset (i.e. task\#5). Taking a look at the whole incremental learning process, we can find that LIQA obtains more stable performance compared with other lifelong learning methods and can effectively mitigate the forgetting in the face of the apparent dataset shift (e.g. LIQA reduces the forgetting at task\#2 and task\#5 compared with EWC, online EWC and SI). The performances of each dataset at the final task are listed in Table \ref{tab:crossdataset_lastsession}. From the table we can find that the performances of LIVE, BID, CLIVE and KonIQ-10K outperforms that of other lifelong learning methods. And the overall performance is also comparable with that of joint training. One thing should be noted is that the performance of KADID-10K of fine-tuning is the best among all methods. It is because fine-tuning only focuses on learning the current task and lacks the ability to retain old knowledge. In contrast, LIQA can achive a balance between stability and plasticity and can well preserve the old knowledge while well learning the new task. Specially, we should note that the addition of KADID-10K impairs the performance of previous datasets even if we adopt joint training. It is because that the image number of KADID-10K is far larger than most of the datasets, the unbalanced data distribution of different datasets impairs the optimal performance of each dataset. To verify the  conjecture, we equip the joint training with pesudo replay, which generates equal number of pseudo features for each dataset. By comparing the results of ``JT'' and ``JT+PR'', we can find that  the performance of each dataset is improved by utilizing pseudo features. It further verifies that pseudo features can address the unbalanced problem to some extent by controlling the quality range and the dataset.

\begin{small}
\begin{table}[htbp]
\caption{Performance comparison across various datasets at the final task.  The best performance of each dataset among fine-tuning and the lifelong learning methods is highlighted in bold. JT+PR denotes joint training equipped with pseudo features replaying. }
\label{tab:crossdataset_lastsession}
\setlength{\tabcolsep}{1.5mm}{
\begin{tabular}{c|cccccc}
\hline
 & LIVE & CSIQ & BID & CLIVE & KonIQ-10K & KADID-10K \\ \hline
FT & 0.819 & 0.663 & 0.398 & 0.298 & 0.573 & \textbf{0.942} \\
EWC & 0.829 & 0.623 & 0.416 & 0.362 & 0.647 & 0.922 \\
\begin{tabular}[c]{@{}c@{}}Online \\ EWC\end{tabular} & 0.812 & \textbf{0.771} & 0.530 & 0.446 & 0.657 & 0.916 \\
SI & 0.820 & 0.725 & 0.580 & 0.453 & 0.699 & 0.900 \\
LIQA & \textbf{0.844} & 0.705 & \textbf{0.642} & \textbf{0.572} & \textbf{0.713} & 0.898 \\ \hline
\rowcolor[HTML]{EFEFEF}
JT & 0.886 &  0.937 & 0.663 & 0.581 & 0.778 & 0.921\\
\rowcolor[HTML]{EFEFEF}
JT+PR & 0.929 & 0.962 & 0.690 & 0.735 & 0.808 & 0.937 \\ \hline
\end{tabular}}
\end{table}
\end{small}

\subsection{Analysis and discussions}

\subsubsection{Ablation study}
We conduct experiments to verify the  effectiveness of the Split-and-Merge distillation strategy as well as  the  feature distillation loss and the pseudo replay loss in equation \ref{equ: full_single}. We follow the experimental setting in Section \ref{subsec:distortion_shift} when the incremental step set to 1.   Specially, to verify the effectiveness of the Split-and-Merge distillation strategy, we replace the $L_{PR}$ defined in equation \ref{equ:pr} with:
\begin{equation}
\begin{aligned}
{{L}^{PR}}={{\mathbb{E}}_{z\sim \mathcal{N}(\mathbf{0},\mathbf{1}),s \sim \bar{\mathcal{S}},j \sim p_{j_{t<\mathcal{T}}} }}\left[{{\left\| {{V}_{\mathcal{T}-1}}({{{\tilde{h}}}_{j}})-{{V}_{\mathcal{T}}}({{{\tilde{h}}}_{j}}) \right\|}_{2}}\right],
    \end{aligned}
\end{equation}
where ${{V}_{\mathcal{T}-1}}$ denotes the single-head regression network trained at the former task.

We implement  three variants of LIQA and average the correlation indexes and the forgetting indexs over all  tasks to represent the overall performance during the whole incremental learning process, which is denoted as $\bar{C}$ and $\bar{F}$ respectively. The comparison results of the three variants of LIQA as well as LIQA are shown in Table. \ref{table:ablation}.
From the table we can see that pseudo replay plays a significant role in LIQA. Besides, without the Split-and-Merge strategy, $\bar{C}$ drops and $\bar{F}$ increases. It is due to that the pseudo labels given by the former single-head network are inaccurate when the former task has negative effect on most of the learned distortions (e.g. task\#5). Directly distillating knowledge from the inaccurate  single-head network will lead to the error-propagation problem thus hindering the consolidation of the previous distortions. In contrast, LIQA preserves the response of each previously learned distortion by the auxiliary multi-head regression network and resists the error-propagation problem. The feature distillation loss  constrains that the features extracted by the current feature extractor $V_{\mathcal{T}}$ don't shift away from that by $V_{\mathcal{T}-1}$. The generator replays the pseudo features that resemble the data distribution of the features generated by $V_{\mathcal{T}-1}$ and the regression head is trained with the pseudo features. The drastic change of the features will lead to the mismatch between the real features of the previous distortions and the regression head trained with the pseudo features when testing. As shown in Table \ref{table:ablation}, the $\bar{C}$ of LIQA w/o FD  drops and $\bar{F}$ increases compared with LIQA.

\begin{table}[htbp]
 \centering
\caption{Average correlation index and average forgetting index  of different variants of LIQA. FD represents feature distillation and PR represents pseudo replay.}
\label{table:ablation}
\begin{tabular}{c|ccccc}
\hline
 & w/o Split-and-Merge & w/o FD & w/o PR & LIQA \\ \hline
$\mathbf{\bar{C}}$ & 0.624  & 0.658 & 0.566 & 0.695 \\
$\mathbf{\bar{F}}$ & 0.117  & 0.089  & 0.214 & 0.087 \\ \hline
\end{tabular}
\end{table}

\subsubsection{Memory replay}
We further explore the memory replay strategy and the size of the replay buffer per batch. The average correlation index and the average forgetting index across tasks  are shown in Table \ref{tab:replay}. Suppose that the allocated replay buffer size is denoted as $N_{buf}$ per batch, and the number of the previously seen distortions is denoted as  $M_{pre}$. The re-scaled quality score is within the range of [0,1]. We split the  re-scaled  quality  range  into  five  interval  segments,i.e.  [0,0.2),  [0.2,0.4),  [0.4,0.6),  [0.6,0.8),  [0.8,1.0].

\begin{small}
\begin{table}[htbp]
\caption{Average correlation index and average forgetting index for different replay strategies and different replay buffer sizes. }
\label{tab:replay}
\centering
\setlength{\tabcolsep}{1.5mm}{
\begin{tabular}{c|cccc|cccc}
\hline
 & \multicolumn{4}{c|}{Replay strategy} & \multicolumn{4}{c}{Replay buffer size} \\ \hline
 & Random & Qua & Dist & \begin{tabular}[c]{@{}c@{}}Qua\\ \&Dist\end{tabular} & 350 & 700 & 1400 & 2800 \\ \hline
$\mathbf{\bar{C}}$ & 0.635 & 0.689 & 0.646 & 0.695 & 0.651 & 0.687 & 0.695 & 0.697 \\
$\mathbf{\bar{F}}$ & 0.112 & 0.072 & 0.075 & 0.087 & 0.092  & 0.076 & 0.087 & 0.081 \\ \hline
\end{tabular}}
\end{table}
\end{small}

We design four variants of replay strategy according to the two conditions of the generator: the quality score and the distortion type. The first variant is the random replay, which means that the pseudo features are generated given the random quality score chosen from the range of [0,1] and the random distortion type chosen from the previously learned distortions (corresponding to $Random$ in Table \ref{tab:replay}). The second variant is to control the quality score and allocate $N_{buf}/5$ pseudo features to each of the five interval segments (corresponding to $Qua$ in Table \ref{tab:replay}). The distortion type is randomly chosen from the previously seen distortions. The third variant is to control the distortion type and allocate $N_{buf}/M_{pre}$ pseudo features to each of the seen distortions (corresponding to $Dist$ in Table \ref{tab:replay}). The quality score is randomly chosen from the range of [0,1]. The fourth variant is the method that LIQA adopts (corresponding to $Qua\&Dist$ in Table \ref{tab:replay}). We allocate $N_{buf}/M_{pre}/5$ pseudo features to each seen distortion each quality interval segment. From Table \ref{tab:replay} we can see that the performance of $Qua\&Dist$ outperforms that of $Random$, $Qua$ and $Dist$. It demonstrates that by well controlling the distortion type and the quality range of the generated pseudo features, LIQA can avoid the unbalanced distribution and enhance the consolidation of the knowledge of learned distortions.

Moreover, we explore the effect of the replay buffer size per batch $N_{buf}$. As shown in Table \ref{tab:replay}, as the $N_{buf}$ increases, the performance improves. Considering the balance between the performance and the complexity, we set $N_{buf}$ to 1400 in LIQA.


\subsubsection{Robustness to task permutations}
To verify the robustness to different task permutations of LIQA, we conduct experiments  with repsect to distortion shift.  We set the incremental step to 1 and randomly permute the order of the 18 novel distortions. We compute the average correlation index and the average forgetting index across all tasks  and the results are shown in Table \ref{tab:permute_distortion}. $Order1$ denotes the default order in Section \ref{subsec:distortion_shift}. From Table \ref{tab:permute_distortion} we can see that the performances of different task permutations vary slightly, which illustrates that LIQA has good robustness to task permutations, benefiting from the Split-and-Merge distillation strategy which resists the negative effect of certain task during the whole incremental learning process.
\begin{table}[htbp]
\centering
\caption{Average correlation index and average forgetting index for different task permutations in terms of distortion shift.}
\label{tab:permute_distortion}
\begin{tabular}{c|ccccll}
\hline
 & Order1 & Order2 & Order3 & Order4 & \multicolumn{1}{c}{Order5}  \\ \hline
$\mathbf{\bar{C}}$ & 0.695 & 0.674 & 0.698  & 0.694 & 0.710   \\
$\mathbf{\bar{F}}$ & 0.087 & 0.070 & 0.082  & 0.113 & 0.086 \\ \hline
\end{tabular}
\end{table}





\subsection{Visualization}
\begin{figure}[htbp]
\centering
\includegraphics[scale=0.47]{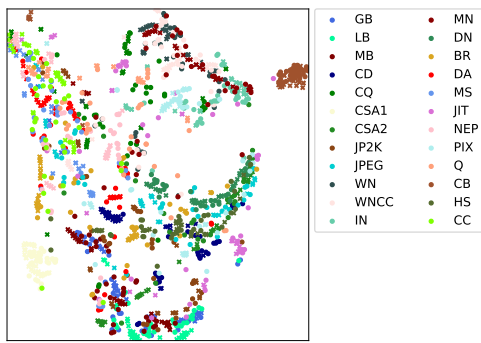}
\caption{t-SNE visualization of 24 previously learned distortions at the final task with incremental set to 1. $\bullet$ corresponds to real features and $\times$ corresponds to pseudo features.}
\label{fig:distortion_vis}
\end{figure}

We visualize the generated pseudo features and the real features at the final task using t-SNE \cite{t-SNE}.
For the distortion shift, we use the feature extractor obtained at the final task to extract some randomly sampled testing images of each previously seen distortion and use the generator trained at the penultimate task  to generate the corresponding pseudo features. The t-SNE visualization is shown in Fig. \ref{fig:distortion_vis}.
We can see that the pseudo features are near the real features, which helps the consolidation of the learned knowledge in LIQA. Specially, we can find that the features of CB is far away from that of the most distortions, explaining why directly fine-tuning on CB will lead to a large degree of forgetting in Fig. \ref{fig:7base_1stepF_order1}.

\begin{figure}[t]
\centering
\includegraphics[scale=0.4]{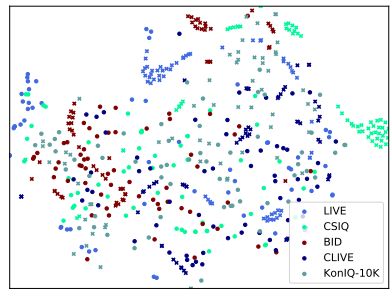}
\caption{t-SNE visualization of 5 previously learned datasets at the final task with incremental set to 1. $\bullet$ corresponds to real features and $\times$ corresponds to pseudo features.}
\label{fig:dataset_vis}
\end{figure}

For the dataset shift, when the incremental step is set to 1, the number of the previously seen datasets is 5. We use the feature extractor obtained at the final task to extract some randomly sampled testing images of each previously seen dataset and use the generator trained at the penultimate task  to generate the corresponding pseudo features. The t-SNE visualization is shown in Fig. \ref{fig:dataset_vis}. From the figure, we can see that compared with the generation of each single distortion in Fig. \ref{fig:distortion_vis}, the generation of each dataset is more difficult due to the low aggregation of data (especially when the image number is limited, e.g. LIVE, CSIQ, BID). However, the requirement of the accurate generation is weakened because we use the auxiliary multi-head regression network to generate pseudo labels for the pseudo features instead of the hard label (i.e. the input quality condition of the generator). Even if the generated pseudo features are not so accurate, they are still near the real features and can help the distillation of the knowledge of the multi-head regression network to build the final single-head regression network.

\section{Conclusions}
In this paper, we propose a new LIQA framework to achieve the lifelong learning of BIQA. The proposed LIQA employs a generator conditioned on the distortion type and the quality score to generate pseudo features, which serves as a memory replayer when learning new tasks. In order to resist the negative effect of certain task during the whole incremental learning process, we employ an auxiliary multi-head regression network to generate predicted quality score of each seen distortion type. It avoids the conflicts between different distortion types and thus improve the robustness to task permutations. Extensive experiments verify that LIQA can effectively mitigate the catastrophic forgetting when facing with continuous distortion type and even dataset shifts. In the future work, we will further explore more effective methods to generate accurate pseudo features especially when the training data is limited and explore the open-set lifelong learning problem of BIQA.

\bibliographystyle{IEEEtran}
\bibliography{references}

\end{document}